\documentclass[10pt,letterpaper]{article}
\usepackage[top=0.85in,left=2.75in,footskip=0.75in]{geometry}

\usepackage{changepage}

\usepackage[utf8]{inputenc}

\usepackage{textcomp,marvosym}

\usepackage{fixltx2e}

\usepackage{amsmath,amssymb}

\usepackage{cite}

\usepackage{nameref,hyperref}

\usepackage[right]{lineno}

\usepackage{microtype}
\DisableLigatures[f]{encoding = *, family = * }

\usepackage{rotating}

\raggedright
\usepackage[aboveskip=1pt,labelfont=bf,labelsep=period,justification=raggedright,singlelinecheck=off]{caption}

\makeatletter
\renewcommand{\@biblabel}[1]{\quad#1.}
\makeatother

\date{}

\usepackage{lastpage,fancyhdr,graphicx}
\usepackage{epstopdf}
\fancyheadoffset[L]{2.25in}
\fancyfootoffset[L]{2.25in}

\begin{document}
\vspace*{0.35in}

\begin{flushleft}
{\Large
\textbf\newline{The limitations of model-based experimental design and parameter estimation in sloppy systems}
}
\newline
\\
Andrew White\textsuperscript{1},
Malachi Tolman\textsuperscript{1},
Howard D.~Thames\textsuperscript{2,3},
Hubert Rodney Withers\textsuperscript{3},
Kathy A.~Mason\textsuperscript{3},
Mark K.~Transtrum\textsuperscript{1,*}
\\
\bigskip
\bf{1} Department of Physics \& Astronomy, Brigham Young University, Provo, Utah, USA
\\
\bf{2} Department of Biostatistics, UT MD Anderson Cancer Center, Houston, Texas, USA
\\
\bf{3} Department of Experimental Radiation Oncology, UT MD Anderson Cancer Center, Houston, Texas, USA
\\
\bigskip

%
%





* mktranstrum@byu.edu

\end{flushleft}
\section*{Abstract}
We explore the relationship among experimental design, parameter estimation, and systematic error in sloppy models.  We show that the approximate nature of mathematical models poses challenges for experimental design in sloppy models.  In many models of complex biological processes it is unknown what are the relevant physical mechanisms that must be included to explain system behaviors.  As a consequence, models are often overly complex, with many practically unidentifiable parameters.  Furthermore, which mechanisms are relevant/irrelevant vary among experiments.  By selecting complementary experiments, experimental design may inadvertently make details that were ommitted from the model become relevant.  When this occurs, the model will have a large systematic error and fail to give a good fit to the data.  We use a simple hyper-model of model error to quantify a model's discrepancy and apply it to two models of complex biological processes (EGFR signaling and DNA repair) with optimally selected experiments.  We find that although parameters may be accurately estimated, the discrepancy in the model renders it less predictive than it was in the sloppy regime where systematic error is small.  We introduce the concept of a \emph{sloppy system}--a sequence of models of increasing complexity that become sloppy in the limit of microscopic accuracy.  We explore the limits of accurate parameter estimation in sloppy systems and argue that identifying underlying mechanisms controlling system behavior is better approached by considering a hierarchy of models of varying detail rather than focusing on parameter estimation in a single model.

\section*{Author Summary}
Sloppy models are often unidentifiable, i.e., characterized by many parameters that are poorly constrained by experimental data.  Many models of complex biological systems are sloppy, which has prompted considerable debate about the identifiability of parameters and methods of selecting optimal experiments to infer parameter values.  We explore how the approximate nature of models affects the prospect for accurate parameter estimates and model predictivity in sloppy models when using optimal experimental design.  We find that sloppy models may no longer give a good fit to data generated from ``optimal'' experiments.  In this case, the model has much less predictive power than it did before optimal experimental selection.  We use a simple hyper-model of model error to quantify the model's discrepancy from the physical system and discuss the potential limits of accurate parameter estimation in sloppy systems.


\section*{Introduction}

Mathematical models play an important role in understanding complex biological systems.  Mathematical models often synthesize a large amount of information about a system into a single representation that can be used to give both conceptual insights into mechanisms and to make predictions about different experimental conditions.  As our mechanistic understanding of the underlying biological processes grows, so too do the scope and complexity of mathematical models used to describe them.  However, mathematical models are never a complete representation of a biological system.  This is a strength, not a weakness, of mathematical modeling.  Mathematical models always include simplifying approximations and abstractions that provide insights into which components of the system are ultimately responsible for a particular behavior\cite{rosenblueth1945role}.  Mathematical models, therefore, ought to represent the judicious distillation of the essence of the behavior in question.

Indeed, biological models cover an enormous range of scales and scopes and mechanistic models are usually formulated in terms of the immediately underlying physical components.  Molecular mechanisms, for example, are modeled as dynamic simulations of complex macromolecules, and systems biology models of gene regulation and protein interactions involve ordinary differential equations for the time evolution of chemical kinetics.  Larger scale phenomenon, such as tumor growth or tissue response to radiation treatments involve models that are more removed from fundamental physics, but nevertheless attempt to reflect the effective mechanisms driving a particular behavior.    This approach is appropriate; it is both impractical and theoretically unsatisfying to ``model bulldozers with quarks''\cite{goldenfeld1999simple}.

Unfortunately, it is very difficult to identify a priori which components of a complex system can be ignored, i.e., which degrees of freedom are \emph{irrelevant}.  It is therefore common for mathematical models to be very complex and include more mechanisms than are strictly necessary to explain a phenomenon.  When overly complex models are fit to data, the parameters associated with the irrelevant mechanisms are difficult to infer from observations.  These parameters are said to be (practically) unidentifiable.  Parameter identifiability is (locally) measured by the Fisher Information Matrix (FIM)\cite{rothenberg1971identification,cobelli1980parameter,chis2014sloppy}:
\begin{equation}
  \label{eq:FIMgeneral}
  I_{\mu\nu} = -\left\langle \frac{\partial^2 \log P(\xi \vert \theta)}{\partial \theta_\mu \partial \theta_\nu}  \right\rangle = \left\langle \frac{\partial \log P(\xi \vert \theta) }{\partial \theta_\mu} \frac{\partial \log P(\xi \vert \theta) }{\partial \theta_\nu} \right\rangle,
\end{equation}
where $P(\xi\vert\theta)$ is the probability distribution for random variable $\xi$ given parameters $\theta$ and $\langle \cdot \rangle$ means expectation value.  (Both $\xi$ and $\theta$ can be vector quantities.)  A small eigenvalue of the FIM indicates that a combination of parameters (given by the corresponding eigenvector) can vary by a large amount without affecting the behavior of the system.  

In many cases, particularly in the context of modeling chemical kinetics in systems biology, the eigenvalues of a model's FIM are ``sloppy'', i.e., have a uniform spacing of FIM eigenvalues on a log scale spread over many orders of magnitude\cite{brown2003statistical,brown2004statistical,waterfall2006sloppy,gutenkunst2007universally,daniels2008sloppiness}.  However, the phenomenon is not unique to biochemical kinetics models; sloppiness has been observed in a wide variety of models in biology, physics, and engineering\cite{frederiksen2004bayesian,waterfall2006sloppy,gutenkunst2007universally,daniels2008sloppiness,machta2013parameter}.  

\begin{figure}
  \includegraphics[width=3.5in]{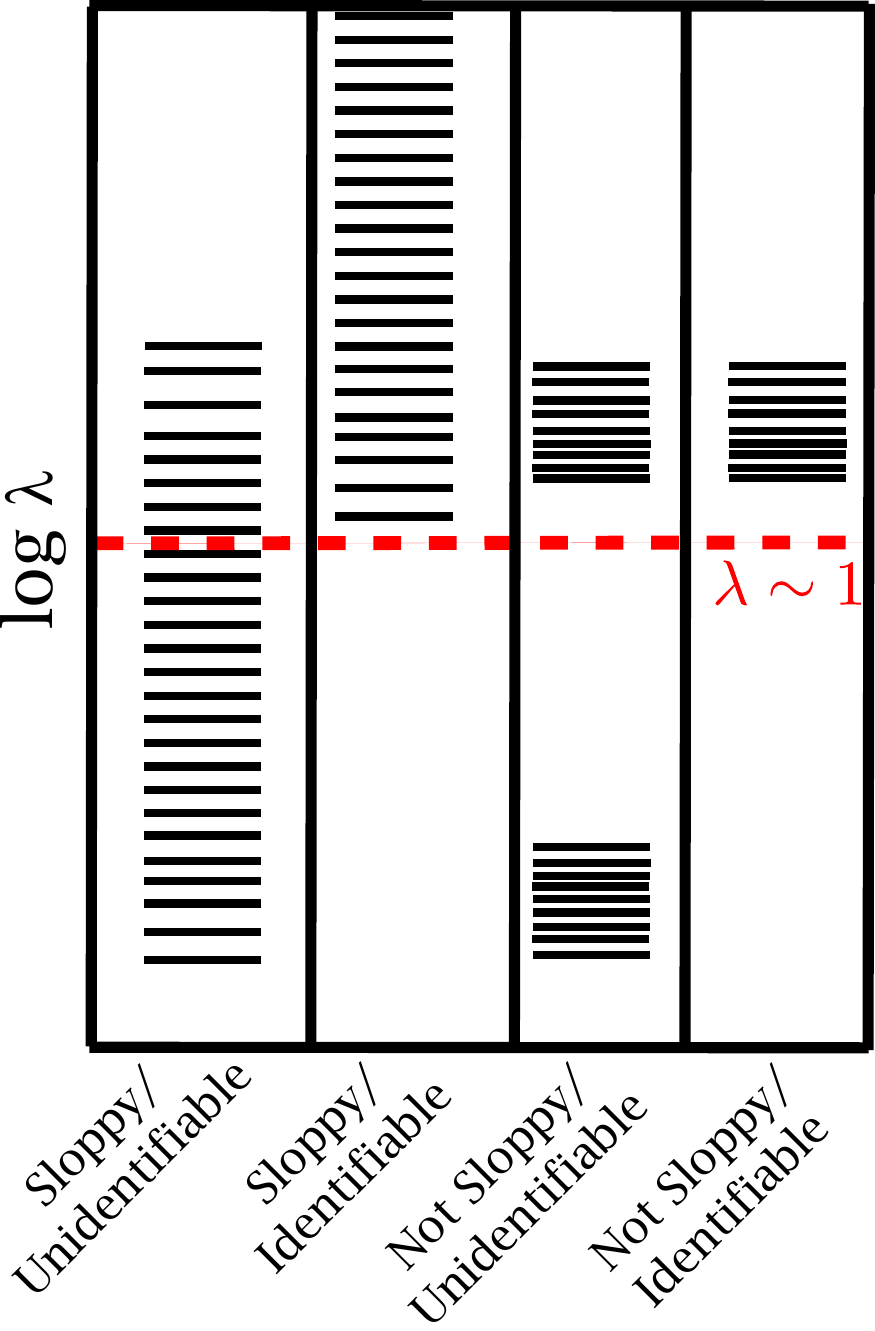}
  \caption{\label{fig:SloppyvsIdentifiable} \textbf{Sloppiness vs. Identifiability.}  Although sloppiness and parameter identifiability are closely related, they are actually two distinct concepts.  Sloppiness refers to an approximate uniform spacing of FIM eigenvalues spread over many orders of magnitude.  In the most commmon case (first column) this means that many eigenvalues will be small and also correspond to unidentifiable parameter combinations.  However, it is possible (in principle) for all the eigenvalues to be large (second column) so that sloppy models can be identifiable (as in references\cite{apgar2010sloppy,chachra2011comment}).  It is also possible for model parameters to be unidentifiable and not sloppy (third column) or identifiable and not sloppy (fourth column).  We here take $\lambda \sim 1$ as the cutoff between identifiable and unidentifiable motived by arguments in Figure~\ref{fig:RelevantDefine}.}
\end{figure}

The exponential eigenvalue distribution of sloppy models quantifies that several parameter combinations are exponentially less important for explaining system behavior than others.  Practically all of the system behavior can be controlled by tuning a small number of stiff parameter combinations (i.e., eigendirections with largest FIM eigenvalues) while varying the sloppy parameter combinations has relatively little effect on the model behavior.  Because of this, sloppiness is closely related to parameter identifiability and is often (incorrectly) used as synonym for practical unidentifiability.  In principle, however, these two concepts are distinct, as we illustrate in Figure~\ref{fig:SloppyvsIdentifiable}.  

Figure~\ref{fig:SloppyvsIdentifiable} illustrates different cases categorized by their sloppiness and identifiability properties.  Sloppy models are characterized by a logarithmic hierarchy of FIM eigenvalues (independent of scale) while unidentifiable models have small eigenvalues (in a sense to be made more precise shortly).  Exampeles of all four possible combinations illustrated in Figure~\ref{fig:SloppyvsIdentifiable} can be found in science.   Examples of sloppy, unidentifiable models (first column) abound in the systems biology literature (see references\cite{waterfall2006sloppy,gutenkunst2007universally,machta2013parameter} for several examples).  Because of the ubiquity of sloppiness in the systems biology, it was suggested that accurate parameter estimation in sloppy models was impossible (or at least impractical) because of the unreasonable data requirements\cite{gutenkunst2007universally}.   However, there are cases of identifiable sloppy models (second column, see for example references\cite{apgar2010sloppy,chachra2011comment,tonsing2014cause}).

Unidentifiable models that are not sloppy (third column) are often characterized by a ``small parameter''.  A small parameter is a small dimensionless number appearing in a model that renders certain aspects of the model unimportant.  The canonical example is a system with well-separated time scales\cite{chachra2012structural}.  In this case, the small-parameter is the ratio of time scales, and singular perturbation theory makes explicit the approximation in which the fast dynamics are slaved to the slow variables.  In general, the small parameter separates which mechanisms can be ignored from those that are relevant.  For a sufficiently small parameter, the gap between the identifiable and unidentifiable parameters becomes very large as in the third column.  

Identifiable, non-sloppy models (fourth column) are those in which all parameters are more-or-less equally easily to infer from data.  In our experience, linear least squares models often fall into this category.  However, models from other fields can also fit this description for appropriate observations\cite{tonsing2014cause} or if the model is reduced\cite{transtrum2014model,transtrum2016bridging}.

It is important to note that being unidentifiable does not mean a model is not predictive.  In many cases, models with very large uncertainties in their parameters may nevertheless make falsifiable predictions\cite{gutenkunst2008sloppiness}.  Indeed, it has been argued that the irrelevance of microscopic complications enables effective modeling at different scales without the need to accurately account for all details\cite{transtrum2015perspective}.  It is therefore possible the unidentifiability is necessary for rather than a obstacle to predictive modeling.

Analyses based the eigenvalues of the FIM are limited for several reasons.  First, the FIM is really only meaningful in the asymptotic limit, i.e., the limit of infinite data and identifiable models.  Second, the FIM eigenvalues are dependent on parameterization.  Specifically, re-parameterizing a model with parameters $\theta = f(\phi)$ where $f$ is some (non-singular) function leads to a new FIM related to the first by
\begin{equation}
  \label{eq:FIMtransform}
  I_\phi = \left( \frac{\partial f}{ \partial \phi} \right)^T I_{\theta} \left( \frac{\partial f}{ \partial \phi} \right),
\end{equation}
where $\partial f/\partial \phi$ is the Jacobian matrix of the parameter transformation, $I_\theta$ is the FIM for the $\theta$ parameterization and $I_\phi$ is the FIM for the $\phi$ parameterization.  For an appropriate choice of $f(\phi)$, it is possible to transform eigenvalues of $I_\phi$ to be any positive numbers.  A simple example of a possible reparameterization is changing the units of parameters (e.g., from meters to nanometers), although in principle any nonsingular $f$ is permissible. 

\begin{figure}
  \includegraphics[width=5.5in]{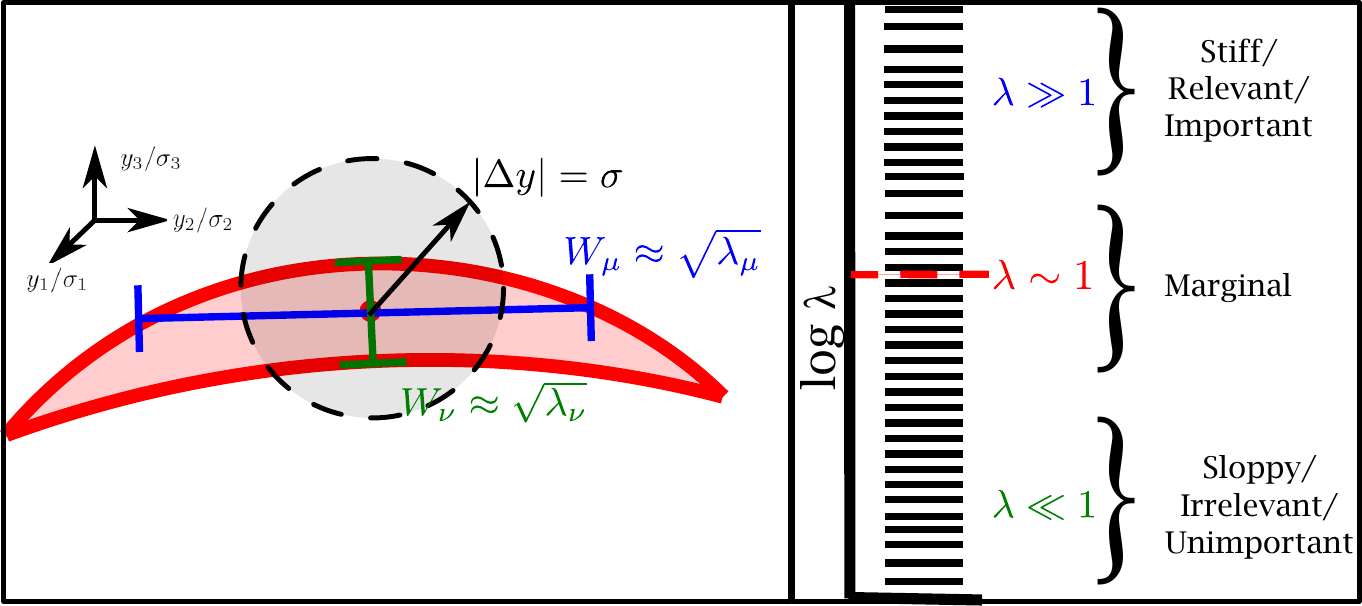}
  \caption{\label{fig:RelevantDefine} \textbf{Model Manifold Widths define relevant and irrelevant parameters.}  (Left) The set of all possible model outputs defines a manifold of predictions.  The true model ideally corresponds to a point near the manifold (red dot).  For typical sloppy models, the manifold is bounded by a hierarchy of widths that are approximately given by the square-roots of the FIM eigenvalues (when parameterized in natural units).  Widths of the model manifold are measured in units of the standard-deviation of the data, so that widths much less than one are practically indistinguishable from noise.  Widths larger than one, on the other hand, are distinguishable from noise and must be tuned to reproduce the observations.  This suggests describing parameter combinations corresponding to large eigenvalues and large widths as relevant or important for the model.  In contrast, those parameters corresponding to small eigenvalues and widths are irrelevant or unimportant.  We describe widths comparable to the experimental noise as marginal.}
\end{figure}

It was shown in reference\cite{transtrum2010nonlinear} using information geometry\cite{murray1993differential,amari2007methods,transtrum2011geometry} that the FIM eigenvalues often reflect a global, parameterization-independent property of the model.  In this approach, the set of all possible model outputs (found by varying the parameters over all physically possible values) generates a manifold of possible predictions with the FIM acting as a Riemannian metric.  This idea is illustrated in Figure~\ref{fig:RelevantDefine}.  For sloppy models, this manifold is often bounded by a hierarchy of widths, reminiscent of the hierarchy of sloppy eigenvalues.  Indeed, when the model is parameterized by dimensionless parameters (e.g., using log-parameters), the widths are approximately given by the square root of the eigenvalues $W_\mu \approx \sqrt{\lambda_\mu}$, as in reference\cite[Figure 3]{transtrum2010nonlinear}.

The concept of manifold widths makes the notion of identifiability distinct from issues related to model parameterization.  For example, a small FIM could be an artifact of a poorly chosen parameterization.  The value of the parameters could indeed vary over a numerically large range, but the apparent unidentifiability may just be a consequence of poorly chosen units (for example).   If this is the case, sufficiently large variations in the parameters will change the model predictions in a statistically significant way.  In contrast, the existence of manifold widths demonstrates that there are parameter combinations that can vary \emph{infinitely} without appreciably affecting the model behavior.  These parameter combinations are truly unidentifiable.

We describe parameters that are unidentifiable as being irrelevant or unimportant. These parameters correspond to manifold widths much less than the scale of the experimental noise, i.e., FIM eigenvalues much less than one.  These parameter combinations could be fixed to arbitrary values or removed from the model without affecting the ability of the model to give a good fit to the data.  In contrast, parameters with widths much greater than the experimental noise need to be tuned to reproduce the observed behavior.  These parameters can be accurately inferred from data, and we call them relevant or important.  Parameters between these two extremes we call marginal.  

Optimal experimental design has been proposed as a way to improve the identifiability of model parameters\cite{pukelsheim1993optimal,faller2003simulation,cho2003experimental,casey2007optimal,balsa2008computational,apgar2008stimulus,apgar2010sloppy,erguler2011practical,chachra2011comment,transtrum2012optimal,chung2012experimental}.  Experimental design is a broad subject; in this paper, we use the term to refer to the general process of using numerical simulations to identify potential experiments to render model parameter identifiable.  Many different approaches are available\cite{mdluli2015efficient,he2010maximin,bazil2012global,vanlier2012bayesian,weber2012trajectory,busetto2013near,liepe2013maximizing,pauwels2014bayesian}.  Consider, for example, the model of Brown et al.\cite{brown2003statistical} of EGFR signaling.  This model has 48 unknown parameters (mostly reaction rates and Michaelis constants) and is the seminal example of a sloppy model\cite{brown2003statistical,brown2004statistical,gutenkunst2007universally}.  Apgar et al.\cite{apgar2010sloppy} identified five experiments (from a candidate pool of 164,000) for which, if performed, the FIM would have no small eigenvalues and all parameters could be identifiable.  Notably, the model was unidentifiable for each of the experiments individually; however, the unidentifiable parameter combinations were different for each experiment.  When the experiments were fit collectively, therefore, the parameters could all be estimated to within 10\%, so the experiments were described as \emph{complementary}\cite[Figure 1]{apgar2010sloppy}.  Although these optimal experiments still require considerably more data than is typical in order to have the desired effect\cite{chachra2011comment}, the dramatic reduction in the range of FIM eigenvalues (and manifold widths) is nevertheless encouraging.  Subsequent work\cite{chis2014sloppy,tonsing2014cause} has confirmed this result: carefully chosen, complementary experiments can have a dramatic effect on parameter identifiability.  When selecting optimal experiments, it is of course important to limit the search to those experiments for which the model is valid.  However, for in many cases it is difficult to know a priori which experiments these will be.

In practice the true distribution (red dot) in Figure~\ref{fig:RelevantDefine} does not lie on the model manifold.  There are always some systematic errors that result from approximations in the model.  However, if the model contains all of the relevant parameters, the systematic errors will not be larger than the experimental noise and can primarily be ignored.  By including additional mechanisms in the model, the model manifold will come closer to the true distribution, usually at the cost of including additional irrelevant parameters.  We use the terms ``model discrepancy''  and ``model error'' to mean the degree to which the true distribution is realizable by the model.  

In this paper we consider the effects of model discrepancy on parameter estimation in sloppy models, i.e., what effect the approximations inherent to the model have on the prospect of accurate parameter estimation.  Although all mathematical models employ some simplifying approximations, in most well-understood examples, the validity of the approximation can be traced to the existence of a small parameter in the system as discussed above. In addition to suppressing irrelevant details, the small parameter also helps identify under which experimental regime the approximation is valid (those experiments for which the parameter is small).  For example, thermodynamics becomes exact in the limit of large system size.  As long as experimental probes are restricted to sufficiently large systems, the corrections due to statistical mechanical fluctuations will be small and the approximate model can be treated, for all intents and purposes, as being an exact surrogate of the physical reality.  

\begin{figure}
  \includegraphics[width=5.5in]{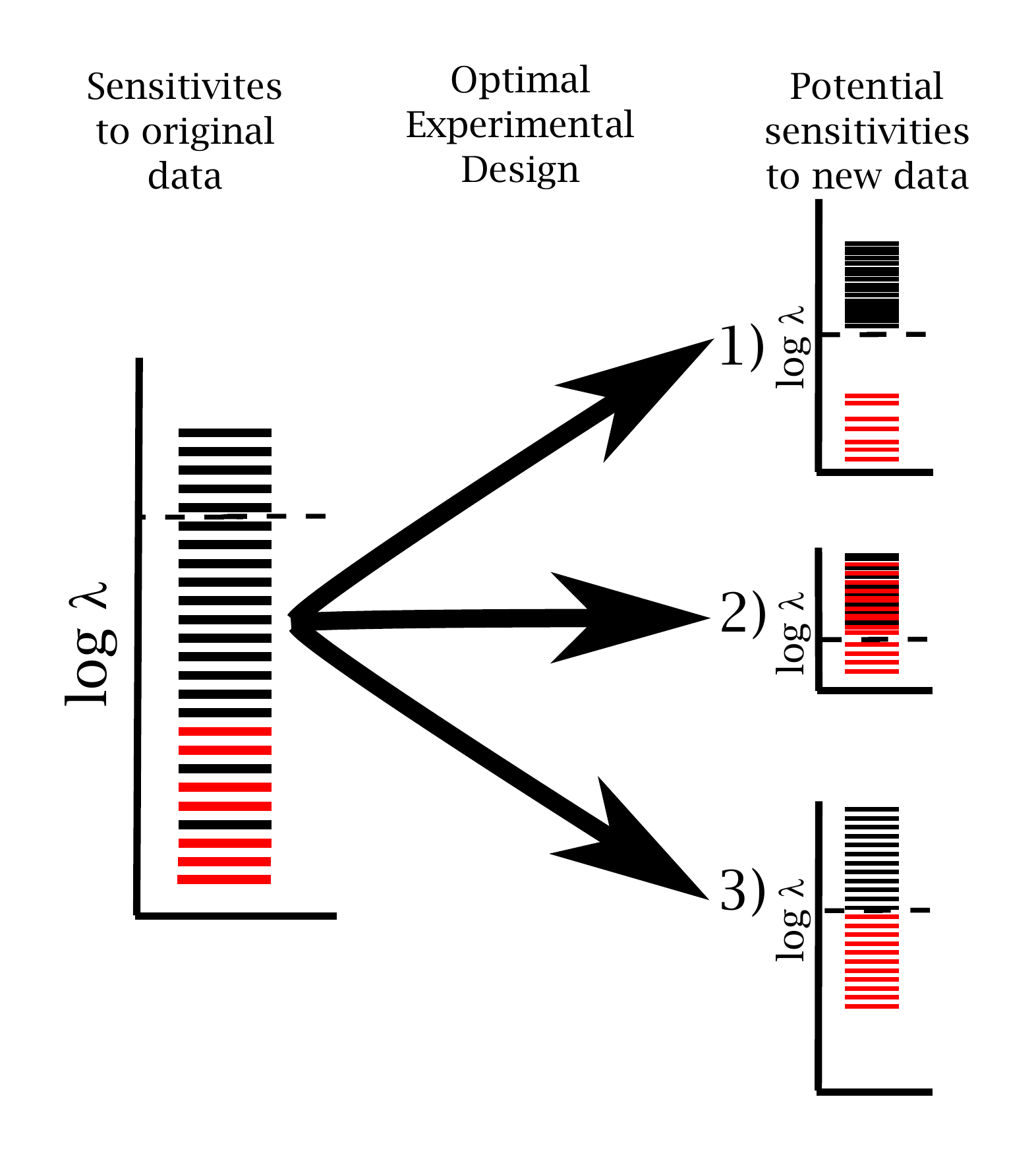}
  \caption{\label{fig:sloppysystemcartoon} \textbf{Experimental Design in Sloppy System.} Sloppy models are characterized by an exponential distribution of FIM eigenvalues (left).  Black lines are FIM eigenvalues for the model in question.  Red lines represent additional eigenvalues that would be introduced by using a more realistic model.  Optimal experimental design selects experiments so as to shift all the black eigenvalues above some desired threshold (dashed line).  Under these new experimental conditions, the red eigenvalues could (1) remain irrelevant, (2) become relevant, or (3) become marginally relevant.}
\end{figure}

However, in many physical systems there is no obvious small parameter, making it difficult to know a priori which physical details are relevant\cite{machta2013parameter}.  (This is one reason that models often include unnecessary details.)  What is more problematic, the near-uniform spacing of FIM eigenvalues in sloppy models suggests that there is no clear separation between relevant and irrelevant details.  If more mechanisms were added to a sloppy model, there would be more FIM eigenvalues.  However, this new model will likely be sloppy too, so that the new eigenvalues will not be much smaller (on a log scale) than the those of the original model.  Optimal experimental design chooses complementary experiments so as to make the small eigenvalues become larger.  What then is the effect of this process on the eigenvalues of the more accurate model?  Do they also become larger?  If so, the approximate model will not be able to fit the data accurately, leading to less predictive models.  This question is illustrated in Figure~\ref{fig:sloppysystemcartoon} and the primary purpose of this paper is to explore this possibility.  

In this paper we demonstrate the complicated relationship between models and experimental conditions using two models of EGFR signaling.  We then introduce a simple hyper-model to quantify the systematic error in the model.  We next consider models of DNA repair as a second example.  We find in both cases that the model's predictive power is appreciable reduced after fitting to optimally chosen experiments.  This loss in predictive power is the result of larger systematic errors which come in spite of more accurately constrained parameter values.

\section*{Results}

\subsection*{ EGFR Signaling}

We begin with an illustrative example; consider a model of EGFR signaling due to Brown et al.\cite{brown2003statistical}.  The model contains 48 unknown parameters (reaction rates and Michaelis constants) that were originally fit to 63 data points\cite{brown2003statistical}.  These data gave limited measurements of activity levels a few proteins after EGF and NGF stimulation in conjunction with a few knockout perturbations.  Notably, this model gave a reasonable fit to experimental (not simulated) data.  Furthermore, the model also made accurate, falsifiable predictions for the system behavior under novel perturbation experiments that were subsequently verified.  From this, we conclude that the model likely contained all of relevant biology and chemistry necessary for giving a mechanistic explanation of these observations.  However, the parameters were largely unconstrained when fit to this data, with relative uncertainties ranging from a factor of 50 up to a factor of a million.

In order to constrain the parameter estimates, Apgar et al.\cite{apgar2010sloppy} proposed a set of five experimental conditions specifically for the model of Brown et al..  These experiments were selected from a candidate pool of more than 160,000 possible experiments that included EGF and NGF stimulations at various levels and in various combinations, as well as potential knockout and over-expression perturbations.  The five optimal experiments were chosen according to a ``greedy'' algorithm to maximize the smallest eigenvalue of the FIM.  Under these experimental conditions, the authors expected to estimate all parameters to better than 10\% accuracy.  The experiments were simulated and not physically performed.  

Rather than actually perform the five experiments proposed by Apgar et al., we create a second model to act as a surrogate for reality.  The EGFR system is well studied and a model that reflects our current understanding of the system\cite{oda2005comprehensive} would contain far more than the 48 parameters of Brown et al..  As a reasonable next step along the ladder of realism, we replace the Michaelis-Menten approximations of the Brown et al.~model by the mechanistic model from which this approximation was derived.  The mechanistic model of (Henri-)Michaelis-Menten is two step enzyme-catalyzed reaction obeying mass-action kinetics: $E + S \rightleftharpoons ES \rightarrow E + P$.  This change introduces several new parameters (from 48 to 70) as well as several new chemical species corresponding to the intermediate enzyme-substrate complexes. 

Note that there are a large class of mechanisms that could result in the same approximate Michaelis-Menten equation, so implementing the mechanistic model described above also makes several other simplifying assumptions.  For example, a mechanistic model could also be written as  $E + S \rightleftharpoons ES \rightleftharpoons EP \rightarrow E + P$, i.e., with an isomerisation step for the enzyme-substrate complex.  Indeed, there is a hierarchy of refining approximations one could make to this model.  Our choice represents what is likely the simplest next step in refining the mechanistic description of Brown et al..   For brevity, we refer to the original model of Brown et al.~that implements the Michaelis-Menten approximation as the approximate model.  We refer to the model of the Michaelis-Menten mechanism with mass-action kinetics as the mechanistic model.  

\begin{figure}
  \includegraphics[width=5.5in]{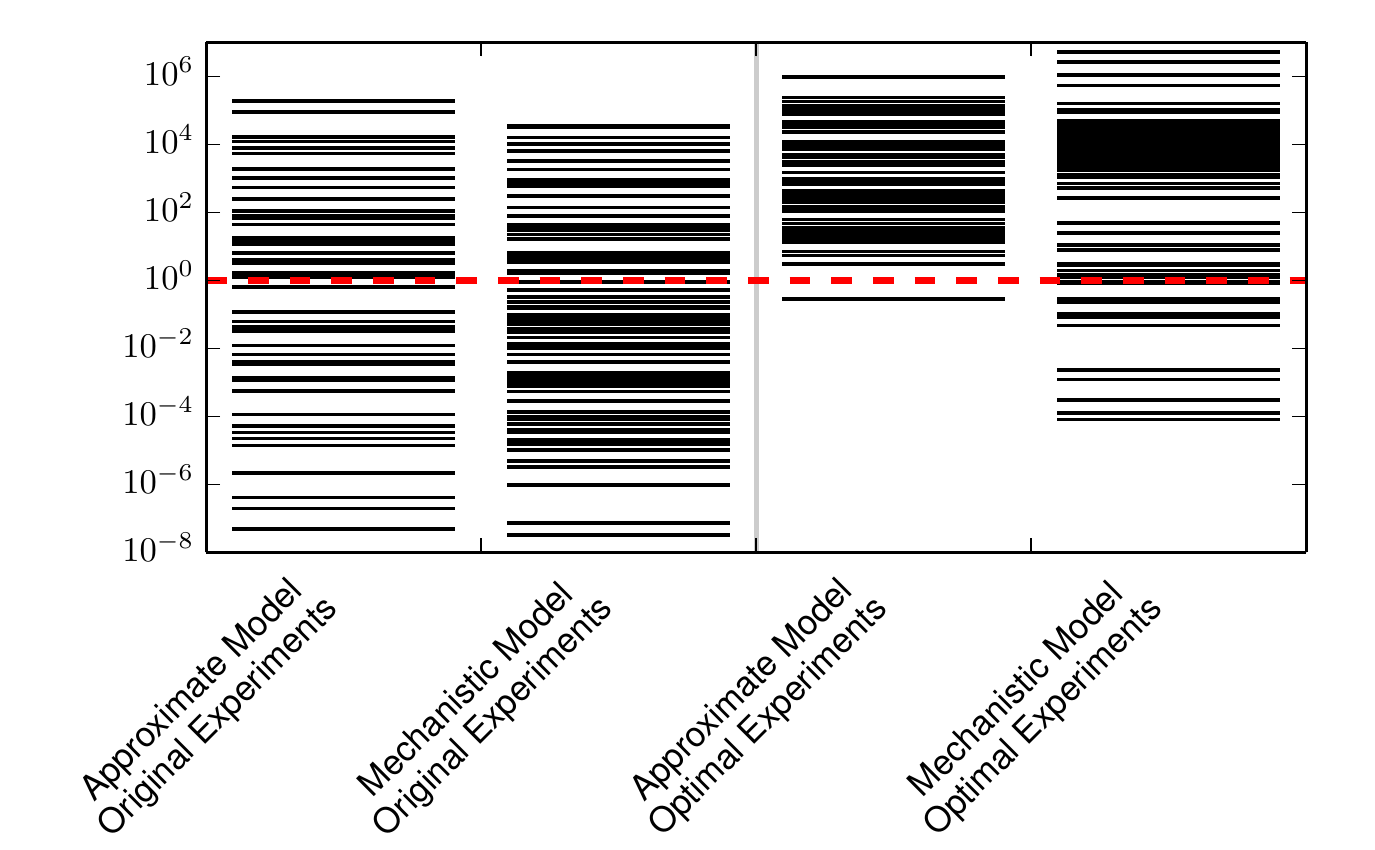}
  \caption{\label{fig:eigenvaluesegfr} \textbf{FIM for the four EGFR models.} Both the approximate Michaelis-Menten kinetics and mechanistic mass-action kinetics are unidentifiable when fit to the data in reference\cite{brown2004statistical}.  Although the optimal experiments in reference\cite{apgar2010sloppy} lead to an identifiable (but still sloppy) model for the approximate Michaelis-Menten kinetics, the mechanistic mass-action kinetics remain unidentifiable.  Furthermore, the FIM of the mass-action model suggests that a minimal model should include \emph{at least} 60 parameters to explain the expanded observations, i.e., the manifold has approximately 60 widths larger than the experimental noise.  The approximate Michaelis-Menten kinetics do not contain all of the relevant physics. The red dashed line corresonds to a relative standard error of $1/e$ in the inferred parameters.}
\end{figure}

Using the model and parameter values of Brown et al., we simulate the experimental conditions from reference\cite{brown2003statistical} and add random noise to generate an initial data set characteristic of that in reference\cite{brown2004statistical}.  We then fit the 70 parameter, mechanistic EGFR model to this initial data.  The resulting fit is both sloppy and unidentifiable as can be seen in Figure~\ref{fig:eigenvaluesegfr} (second column).  Notably, the 70 parameter model has 22 more eigenvalues than the 48 parameter model, but there is not a clear separation between the largest 48 and the smallest 22 eigenvalues.  Furthermore, it is not possible to equate the largest 48 eigenvalues with the 48 parameters of the approximate model and the 22 smallest eigenvalues with the new parameter combinations introduced by the more detailed kinetics.

Next, we simulate artificial data for each of the experiments suggested by Apgar et al. using the 70 parameter mechanistic model and parameter values estimated from the first fit.  We add random noise to these simulations to create a second data set.  This second data set, having come from the more complicated model, acts as a surrogate for real experimental data.  We then fit the 48 parameter, approximate model to the second artificial data.  

The use of Michaelis-Menten kinetics in models of protein networks is somewhat controversial.  In many practical cases, such as the original experiments of Brown et al., the model seems to work.  However, the actual condition that must be satified is shown rigorously in reference\cite{segel1989quasi} to be
\begin{equation}
  \label{eq:MMcond}
  \frac{[E]}{[S] + K_M} \ll 1,
\end{equation}
where $[E]$ and $[S]$ are the enzyme and substrate concentrations respecively and  $K_M = (k_r + k_{cat})/k_f$ is the Michaelis constant.  From this condition we see that it is not strictly necessary for $[E] \ll [S]$ as is often (incorrectly) asserted.  However,  this condition is derived for a single enzyme-substrate reaction in isolation and, as we will see below, is generally not a sufficient condition for the Michaelis-Menten approximation to hold in a network context.  Nevertheless, Eq.~\eqref{eq:MMcond} suggests that the Michaelis-Menten model could be valid provided $K_M$ is very large, even if $E \sim S$.

In choosing parameters for the mechanistic model, we therefore choose parameters such that the combination $K_M= (k_r + k_{cat})/k_f$ is very large.  This is always possible because the model is unidentifiable when fit to the initial data.  We were able to choose parameter values for all reaction rates such that the Michaelis-Menten approximation would not give errors larger than 10\% for any of the reactions given the conditions of the network.  This means that there is no a-priori reason to think that the approximate model of Brown et al. would be a poor approximation to the mechanistic model for our parameter values. 

\begin{figure}
  \includegraphics[width=5.5in]{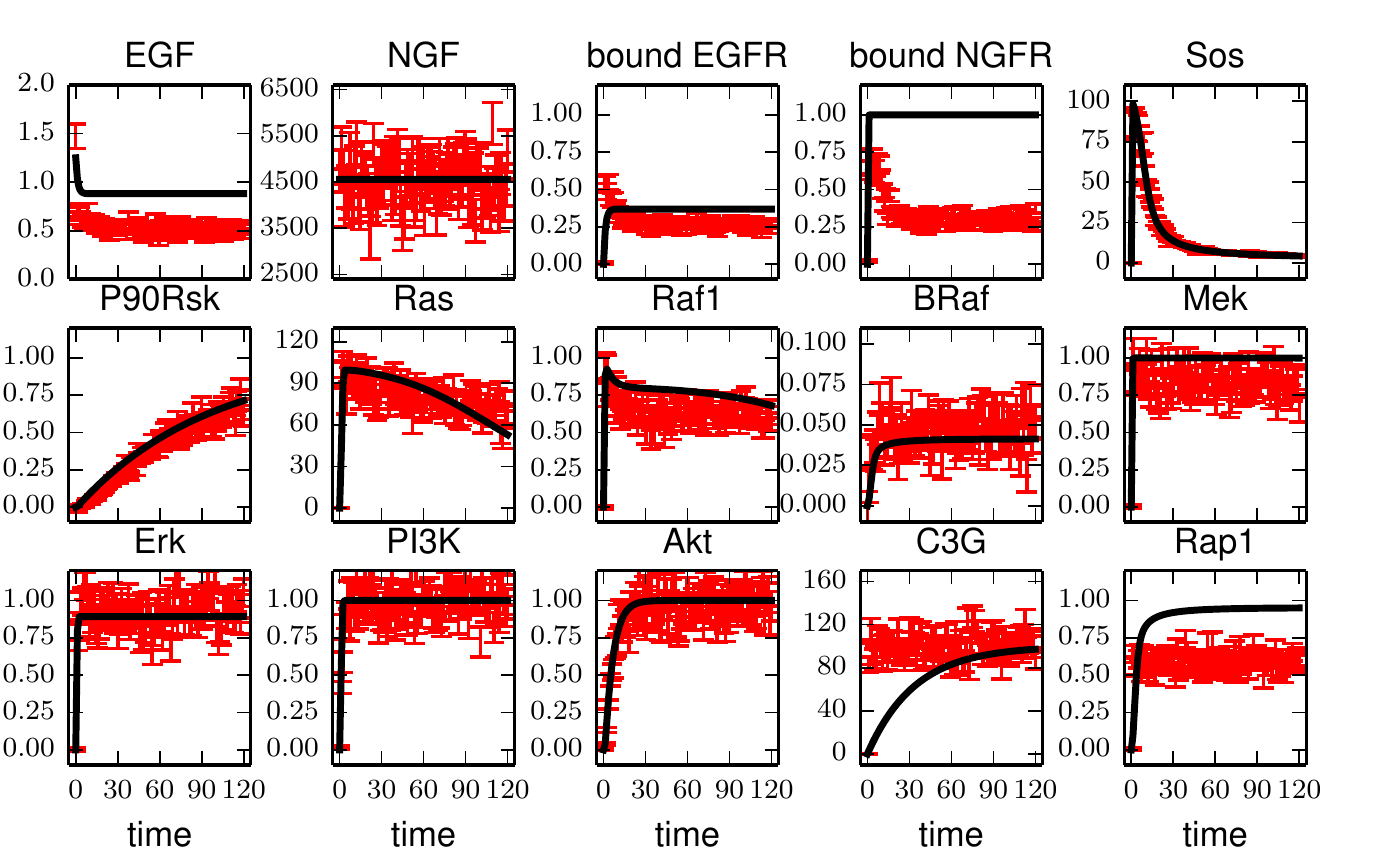}
  \caption{\label{fig:egfrfit} \textbf{Fit of approximate Michaelis-Menten kinetics to mechanistic mass-action data.}  Because the approximate model does not contain all of the relevant mechanisms for the expanded observations, it cannot give a reasonable fit (i.e., within the expected variance of the experimental noise) to all of the experiments simultaneously.  We see here that several time series are fit quite badly which could guide a modeler in identifying the missing relevant mechanisms.}
\end{figure}

Even when enforcing the constraint that $K_M$ is large, we find that the approximate model cannot give a reasonable fit to the data generated by the mechanistic model for the Apgar experiments.  This is illustrated in Figure~\ref{fig:egfrfit} where the systematic errors in the fit are clearly much larger than random noise. 

Because the error in the fit is large, we conclude that approximate Michaelis-Menten kinetics do not contain the relevant mechanisms to explain the observations under the expanded experimental conditions.  This can be seen from the FIM eigenvalues in Figures~\ref{fig:eigenvaluesegfr}.  In particular, observe that the FIM for the mechanistic mass-action model under the expanded experimental condition contains approximately 60 eigenvalues larger than 1, so that there are about 60 directions on the model manifold with widths larger than the experimental noise.  This indicates that a minimal model would require about 60 parameters to fit this data; the 48 parameter model of Brown et al.~is clearly insufficient.   Furthermore, the parameters of the mechanistic mass-action model are unidentifiable for the experimental conditions of Apgar et al..  If the approximate model were replaced by the mechanistic model, it would require another round of experimental design in order to find accurate parameter estimates.

Because the fit of the mechanistic mass action kinetics to the original experiment is unidentifiable, there are many possible parameter values that would give equivalent fits.  We generated an ensemble of parameter values for the mechanistic model consistent with the original experiments.  When these ensembles generate data for the optimal experiments, we consistently find that the best fits for the approximate model have large errors similar to that in Figure~\ref{fig:egfrfit}.

The discrepancy between the data and the fit in Figure~\ref{fig:egfrfit} is reminiscent of over-fitting.  This is not the case however.  Over-fitting occurs when the fit to a training data set is very good (i.e., too good) so that the predictions on a test set suffer as a result.  In this language, the fit in Figure~\ref{fig:egfrfit} is the training set (not the test set), so this is not an instance of over-fitting.  Rather it is a demonstrationg that the model cannot fit the data.

\subsection*{Quantifying Model Error in Sloppy Systems}

Motivated by the example of optimal experimental design in EGFR signaling described above, we now propose a simple method to quantify model descrepancy when fitting data to approximate models.  

We assume that parameters are estimated by least squares regression, although many of our methods and results generalize.  Least squares regression is justified by the assumption that the data are generated from a model with additive Gaussian noise:
\begin{equation}
  \label{eq:data}
  d_i = y_i(\theta) + \sigma_i \xi_i
\end{equation}
where $d_i$ is the i-th data point, $y_i$ is the model prediction for the i-th data point, $\theta$ is a vector of parameters, $\xi_i$ is a random variable with zero mean and unit variance, and $\sigma_i$ is the scale of the noise.  The random variable may account for any intrinsic stochasticity in the system, e.g., thermal fluctuations in particle number or inconsistencies in experimental measurements.  It may also accommodate systematic errors such as model approximations, i.e., mechanisms that have been left out of the model.  The field of uncertainty quantification has begun to explore methods to account for the inadequacy of the model\cite{smith2013uncertainty,kennedy2001bayesian,qian2006building,apley2006understanding,bayarri2007framework,reichert2009analyzing}.  Minimizing the weighted sum of square errors
\begin{equation}
  \label{eq:cost}
  \chi^2(\theta) = \sum_i \frac{(d_i - y_i(\theta))^2}{\sigma_i^2}
\end{equation}
corresponds to a Maximum Likelihood Estimate (MLE) of the parameters for the model in Eq.~\eqref{eq:data}.  

The FIM for the model in Eq.~\eqref{eq:data} is calculated as
\begin{equation}
  \label{eq:FIM}
  I_{\mu\nu} = \sum_i \frac{1}{\sigma_i^2} \frac{\partial y_i}{\partial \theta_\mu} \frac{\partial y_i}{\partial \theta_\nu}.
\end{equation}
The FIM is the inverse covariance matrix for the parameter estimates, so that the square root diagonal elements of the inverse FIM correspond to the one-standard deviation statistical uncertainties in the inferred parameters.  It therefore follows that a well-conditioned FIM is necessary for accurate parameter inference.  



Because mathematical models are always approximations, a model's discrepancy from physical reality can always be improved by including additional physical mechanisms.  This increased realism usually comes at the cost of increased complexity, often in the form of additional parameters, additional physical degrees of freedom, and computational cost.  Therefore, a modeler is often faced with choosing from a hierarchy of models of increasing realism and growing complexity.  Overly complex models increase the possibility of over-fitting data or over-explaining behavior while excessively simple models may lack all the relevant mechanisms.  Effective models strike a balance between these two extremes.

To formalize this concept and facilitate later discussion, we introduce the concept of a \emph{sloppy system}.   We define a sloppy system as a physical system and a set of experimental protocols that can be approximated by a hierarchy of mechanistic, mathematical models of growing complexity that become sloppy in the limit of microscopic accuracy.  

To make this definition more concrete, consider a model of a biological system.   Of necessity, the model does not include every mechanism that is present in the true physical system.    It is therefore possible to augment the model with additional mechanistic details, resulting in a more realistic, but more complex model.  By repeating this process, one can generate a sequence of models of increasing realism that are better approximations to the actual physical system.  The limit of this sequence converges on a model indistinguishable from the physical system, i.e., the limit of microscopic accuracy.

Naturally, this sequence of models will introduce many parameters that are unidentifiable.  However, if the sequence of models is not just unidentifiable, but is also sloppy (i.e., as in the first column of Figure~\ref{fig:SloppyvsIdentifiable} rather than the third column), then we say that the system is sloppy.

As an aside, we have introduced the concepts of ``sloppy systems'' and ``limits of microscopic accuracy'' as useful abstractions.  In practice, constructing  more detailed mechanistic models may have a number of challenges.  For example, it may be necessary to include parameters describing the experimental apparatus but that are separate from the biological system of interest.  These potential complications are beyond the scope of this paper which instead uses these idealized concepts to explain our reported results.

The significance of a sloppy system is that models will always include marginal parameters so that there is always a trade off between model identifiability and model predictivity.  For a fixed set of experimental protocols, there will always be some mechanistic details that are not entirely identifiable but still facilitate predictivity in the model.

The concept of a sloppy system brings together a number of concepts that are each well-known to modelers of complex systems.  Considering multiple models of a single physical system is as old as science, but has recently been employed with new sophistication in the context of ensemble modeling\cite{kuepfer2007ensemble} and multi-scale modeling\cite{feig2004mmtsb,ayton2007multiscale,murtola2009multiscale}.  The literature for parameter identifiability and the related concept of sloppiness was discussed in the introduction.  By bringing these concepts together, we seek to explain the results of our EGFR simulations above and argue that sloppy systems pose unique challenges for predictive modeling.



\begin{figure}
  \includegraphics[width=5.5in]{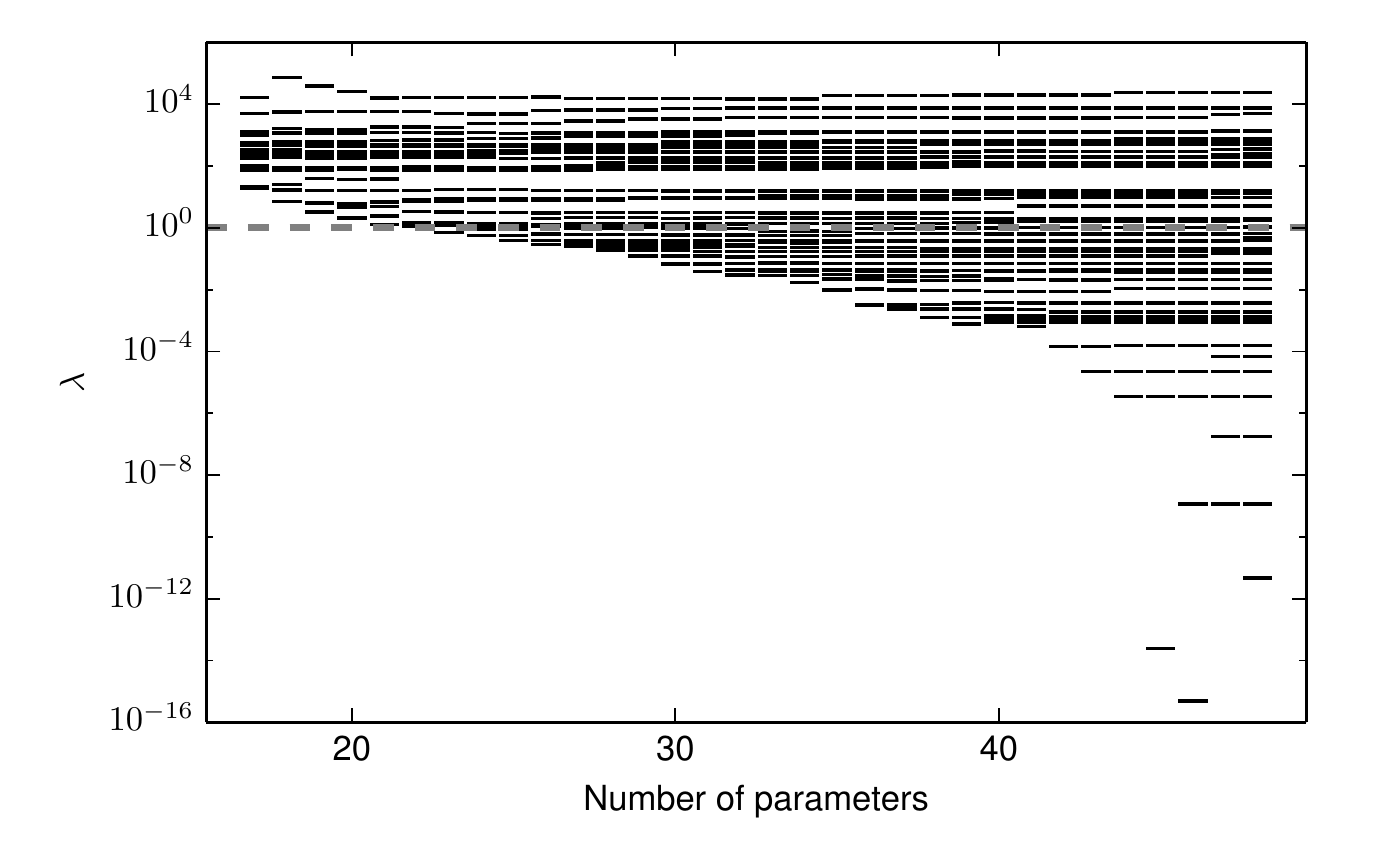}
  \caption{\label{fig:pc12eigenvaluesa} \textbf{An Example of a Sloppy System.} Observations of an EGFR signaling network can be explained by a model that is identifiable and not sloppy. The 18 parameter model has FIM eigenvalues that span fewer than 4 orders of magnitude and are all larger than one.  By including additional mechanisms in the model (more parameters) the models become increasingly sloppy and less identifiable.  The FIM eigenvalues ultimately span more than 16 orders of magnitude, leading to the large parameter uncertainties reported in reference\cite{brown2004statistical}.}
\end{figure}

We now demonstrate that the EGFR system is a sloppy system.  Figure~\ref{fig:pc12eigenvaluesa} shows the FIM eigenvalues for several models of EGFR signaling under the experimental conditions of Brown et al.\cite{brown2003statistical}.  Including more microscopic realism in the model requires additional parameters that render the model less identifiable.  Note that this sequence of models was constructed in reverse--beginning from a sloppy model, irrelevant parameters were removed one at a time using the manifold boundary approximation method\cite{transtrum2014model,transtrum2016bridging}.  Here we reinterpret this result as a demonstration of the existence of sloppy systems and emphasize the trade-off between mechanistic accuracy and model simplicity.  In general, all the models that belong to a sloppy system cannot be ordered in a simple sequence as in Figure~\ref{fig:pc12eigenvaluesa}.  There will often be a complex, hierarchical relationship among models similar to that described by the adjacency graphs in reference\cite{transtrum2014information}.

In order to account for errors due to ignoring marginal parameters, we need to refine the assumptions underlying Eqs.~\eqref{eq:data}-\eqref{eq:FIM}.  If the stochastic term in Eq.~\eqref{eq:data} is dominated by experimental noise, then $\sigma_i$ can be estimated by repeated observations of data point $i$, in which case $\sigma_i$ scales like $\sigma_i \sim 1/\sqrt{n_i}$ where $n_i$ is the number of repeated observations of data $i$.  In this case, $\sigma_i$ becomes the standard deviation of the observations and is a measure of experimental reproducibility.  Henceforth, we assume that $\sigma_i$ denotes the experimental uncertainty and is known from experimental observations.  We modify equation~\eqref{eq:data} to also include an error term $\delta_i$ due to approximations in the model
\begin{equation}
  \label{eq:dataerror}
  d_i = y_i(\theta) + \sigma_i \xi_i + \delta_i.
\end{equation}
The model error term $\delta_i$ will be a type of ``hyper-model'' that accounts for and quantifies errors in the model in a phenomenological way without including additional mechanisms.

\begin{figure}
  \includegraphics[width=5.5in]{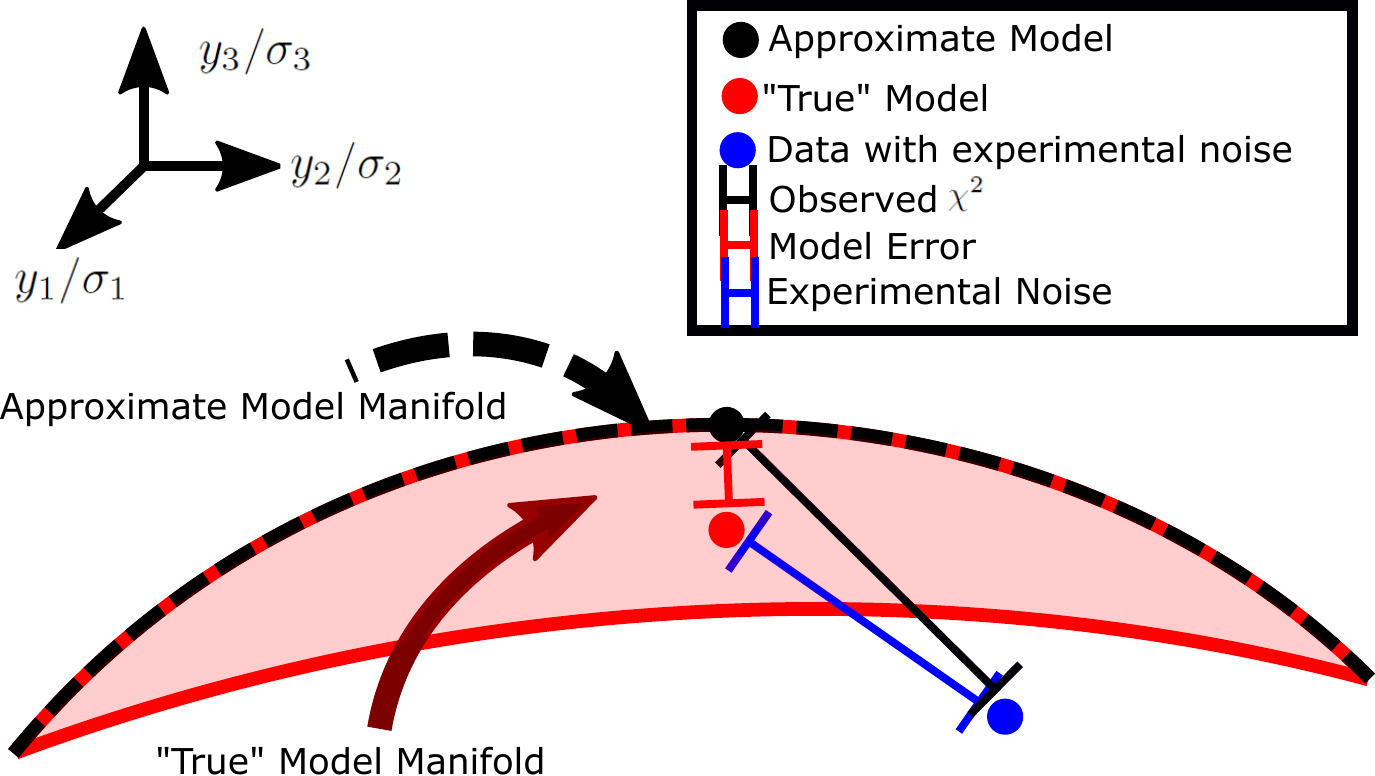}
  \caption{\label{fig:uqmanifold} \textbf{Quantifying Model Error.} As in Figure~\ref{fig:RelevantDefine}, the model of interest forms a statistical manifold in data space, represented by the black dashed line.  Another more realistic model also forms a statistical manifold of higher dimension (red surface).  Experimental observations (blue dot) are generated by adding Gaussian noise of size $\sigma$ to a ``true'' model (red dot).  The least squares estimate is the point on the approximate model (black dot) nearest to the experimental observations.  However, the distance from the best fit to the observed data has contributions from both the experimental noise and the model error.}
\end{figure}

We adopt a simple hyper-model of the systematic error given by
\begin{equation}
  \label{eq:dataerror2}
  \delta_i = f \sigma_i \xi_{i}'
\end{equation}
where $f$ is a hyper-parameter that will be estimated from the data, and $\xi_{i}'$ is another Gaussian random variable with zero mean and standard deviation of one.  We illustrate this concept geometrically in Figure~\ref{fig:uqmanifold}.  When this ansatz breaks down, it is an indication that relevant mechanisms are missing from the model, i.e., that the unfit data has structure that could be modeled and predicted.

Care must be taken in the interpretation of $\xi_i'$.  We have modeled the systematic error as a random variable.  Unlike experimental noise, the size of this uncertainty cannot be decreased by repeated observations.  Rather, the stochastic element in the model error represents the (unknown) approximations in the model.  The relevant statistical ensemble is the set of all possible model refinements that could be made to correct the model shortcomings.

We have assumed that the model errors $\xi_i'$ are uncorrelated among data points.  We also assume that the model is likely to give worse predictions for data points that also have large experimental variation.  These choices are convenient and constitute what is likely the simplest possible such hyper-model.   More sophisticated models could be used, and the meta-problem of modeling the error in the model has been addressed in the context of uncertainty quantification\cite{smith2013uncertainty,kennedy2001bayesian,qian2006building,apley2006understanding,bayarri2007framework,reichert2009analyzing}. In the present context, these assumptions will give us a simple way of estimating the error of the model from data.  These assumptions will be valid provided that $\delta_i$ is small compared to experimental noise.  We will now use our hyper-model to provide a criterion for including additional mechanisms in the model.

The negative log-likelihood for the model of Eqs.~\eqref{eq:dataerror}-\eqref{eq:dataerror2} is
\begin{equation}
  \label{eq:cost2}
  -l(\theta,f) = \sum_i \left( \frac{(d_i - y_i(\theta))^2}{2 \sigma_i^2 \left(1 + f^2 \right)} + \log \sigma_i + \frac{1}{2} \log \left( 1 + f^2\right)+ \frac{1}{2} \log 2 \pi \right).
\end{equation}
The best fit values for the parameters $\theta$ are unchanged by Eqs.~\eqref{eq:dataerror} and \eqref{eq:dataerror2}, and an unbiased estimate of $f$ is given by
\begin{equation}
  \label{eq:festimate}
  f = \sqrt{ \chi^2/(M-N) - 1}
\end{equation}
where $\chi^2$ is the sum of squared error defined in Eq.~\eqref{eq:cost}, $M$ is the number of data points, and $N$ are the number of parameters in the vector $\theta$.  As a practical matter, the model can be fit using Eq.~\eqref{eq:cost} as though there were no approximations in the model.  This is due to our convenient choice of $\delta_i$ in Eq.~\eqref{eq:dataerror2}.

The augmented FIM is given by
\begin{equation}
  \label{eq:FIMthetaf}
  I_{\theta f} = \left(  \begin{array}{cc}
                           \left( \frac{1}{1 + f^2} \right) I & 0 \\
                           0 & 2 \left( \frac{f}{1 + f^2} \right)^2 (M - N)
\end{array} \right),
\end{equation}
where $I$ in the first entry is the $N \times N$ FIM in Eq.~\eqref{eq:FIM}.  The zeros in the off-digonal terms are $1 \times N$ and $N \times 1$ zero vectors.  It follows that the parameter covariance matrix must be modified according to
\begin{equation}
  \label{eq:covariance}
  Cov(\theta) = \frac{\chi^2}{M-N} I^{-1},
\end{equation}
where $I$ is the FIM in Eq.~\eqref{eq:FIM} and $\chi^2$ is the best fit cost that minimizes Eq.~\eqref{eq:cost}.  

It is worth noting that the parameter $f$ contains the same information as the likelihood function as is made explicit by Eq.~\eqref{eq:festimate}.  Indeed, Eq.~\eqref{eq:covariance} is a standard statistical formula for estimating parameter uncertainties in ordinary least squares regression in which the scale of the noise is unknown.

The standard deviation of the estimate in $f$ is given by
\begin{equation}
  \label{eq:deltaf}
  \delta f = \frac{1 + f^2}{f\sqrt{2(M-N)} }.
\end{equation}
We now consider how large an $f$ can be acceptable.  We seek a model's whose approximations do not limit is predictive power.  That is to say, the model error should be small compared to the experimental noise ($f < 1$) and unidentifiable from experimental observations.  If the MLE of model error is $f$, then the statistical uncertainty in that estimate should satisfy $\delta f \sim f < 1$.  If this is the case, then the systematic error will be relatively small and not significantly limit the model's predictive ability.  This criterion gives
\begin{equation}
  \label{eq:fcutoff}
  f  = \frac{1}{\sqrt{\sqrt{2 (M-N)} - 1}}
\end{equation}
as an acceptable value for $f$.


\section*{EGFR Model Revisited}

With this background, we can now revisit the EGFR model above.  The experimental conditions of Apgar et al., included 7000 data points.  If the approximate model were a good approximation, we would expect the fit to have a sum of squares error of approximately $7000\pm84$ where the range is one standard deviation of the Chi-squared distribution.  However, fitting the artificial data typically led to a best fit error greater than 100,000 and was never less than 96,000.  This error corresponds to an estimated value of $f = 3.7$ with $\delta f = 0.03$. 

The statistical uncertainty in the model parameters is larger than expected by a factor $\sqrt{1 + f^2} = 3.8$ for $f = 3.7$.  The optimal experiments were designed to give less than $10\%$ error in parameter estimates, so that modified uncertainty would now be less than $40\%$.  Considering several parameters were initially unknown to a factor of a million, $40\%$ appears to be a significant reduction.  Although the parameter estimates remain somewhat constrained, the predictive power of the model is completely lost.  This is because the effective error bars on the data are also larger by a factor of 3.8.  In our simulations we assumed that the fractional activity levels of all proteins were measured to 10\% accuracy.  With 40\% effective error bars, one standard deviation on either side of the mean covers almost the entire range of possible predictions.

In addition to having a large value for $f$, the ansatz of Eq.~\eqref{eq:dataerror2} breaks down for the fitting the EGFR model.  This is clearly seen by inspecting Figure~\ref{fig:egfrfit}.  We speculate that it may be possible to rescue some of the predictive power of the model by implementing a more sophisticated hyper-model, such as introducing a separate $f$ parameter for each time series or including phenomenological parameters to account for correlations in systematic errors.  However, this possibility is beyond the scope of this work, but has been explored in the uncertainty quantification literature\cite{kennedy2001bayesian,apley2006understanding,smith2013uncertainty}.

\subsection*{DNA Repair Models}

We now consider a model to predict the survival of jejunal crypt clonogens after radiotherapy as in reference\cite{sheu2013use}.  The model used is the linear-quadratic (LQ) model\cite{thames1985incomplete} combined with Curtis' lethal-potentially lethal (LPL)\cite{curtis1986lethal} and Kiefer's repair-saturation (RS)\cite{kiefer1988quantitative} models of post-radiation DNA repair.  The LPL and RS models can be combined into a pair of differential equations
\begin{eqnarray}
  \label{eq:lplrs1}
  \frac{du}{dt} & = & \delta_1 r - \frac{\lambda_1}{1 + \epsilon u} - \lambda_2 u - \lambda_3 u^2 \\
  \label{eq:lplrs2}
  \frac{dv}{dt} & = & \delta_2 r + \lambda_2 u + \lambda_3 u^2
\end{eqnarray}
where $u(t)$ and $v(t)$ quantify the repairable and non repairable lesions.  The parameter $r$ is the dose rate and is known from experimental conditions.  The unknown parameters are $\delta_1r$ and $\delta_2r$ which are the rates of induction of potentially lethal and lethal lesions, and $\lambda_1$, $\lambda_2$, and $\lambda_3$ correspond to rates of repair, fixation, and binary misrepair respectively.  $\epsilon$ is the parameter for repair saturation.  Fixing the parameters $\delta_2 = \lambda_2 = \epsilon = 0$ gives the LPL model.  The RS model is recovered by $\delta_2 = \lambda_3 = 0$.  There are a total of six parameters in the composite model, including $3$ that are absent from the LPL formulation and 2 that are absent in the RS formulation.  

Data from reference\cite{sheu2013use} explored survival rate for split doses.  It reported a statistically significant difference between those cells irradiated initially by a large and then a small dose and vice versa.  The RS model (and by extension the composite six parameter model) gave reasonable fits to the data; however, the uncertainty in the inferred parameters was quite large.  Optimal experimental design was used as an avenue to provide better parameter estimates.

In order to identify the optimal experimental conditions for inferring the parameters in Eqs.~\eqref{eq:lplrs1}-\eqref{eq:lplrs2}, the experimental space was first explored numerically.  Four experimental parameters were varied: the radiation level $r$, the size of each radiation dose, as well as the rest time between doses.  In total, 21,870 experimental conditions were considered.  The FIM for each condition was calculated and nine experimental conditions chosen to augment the original 19.  Seven experiments were repeated to confirm results.  The result of these 35 experiments and the subsequent fit are given in the supplemental material.

The fit to the original 19 experiments gave a value $f = 0.76$ with $\delta f = 0.41$, indicating that the model likely lacks some relevant mechanisms.  Furthermore, the ansatz of Eq.~\eqref{eq:dataerror2} is a good approximation for this model and data set.  Because $f$ is not too big (less than one and within two standard deviations of zero) the model remains predictive in spite of its relative simplicity.    Indeed, the model was able to reproduce and explain the asymmetric response to dose size.  However, the fit to the expanded 35 experiments gave $f = 2.0$ and $\delta f = 0.33$ so that $f$ is now more than 6 standard deviations larger than zero.  

Similar to the simulated case of the EGFR system, the model is unable to give a reasonable fit to the data.  However, the shortcomings of the model are only manifest \emph{after} the observation conditions have been expanded.   Unlike the EGFR system, the DNA repair models results are fit to experiment data (not simulated).  The case of crypt cell survival lacks a comprehensive mechanistic description comparable to what has been done for the EGFR pathway.  From the results of the additional experiments, a simple interpretation of the asymmetry in the first data set is that DNA repair is not monoexponential\cite{ang1992impact}. The higher-order terms would contribute disproportionately when intervals are short (first data set). Overall the so-called “repair halftime” (monoexponential) that fits the first data set best (and gives asymmetry) is shorter than the actual “repair halftime” (a composite of several rates, reflecting a more complicated underlying biochemistry).  However, it is not known what additional mechanisms should be included to fit the new data.  Although beyond the scope of this paper, we speculate that closer inspection of the points of model failure could lead to new insights into DNA repair models.  We further speculate that a hierarchy of models similarly exists for the DNA repair, and our inability to fit the expanded data set indicates that a more complete model would have more than six large eigenvalues in its FIM.

Significantly, the predictive power of the model is decreased dramatically when fit to the expanded data set.  After including the new data, we found that the model was no longer able to predict the significant asymmetry in the dose response.  This is similar to the EGFR case in which the effective error after fitting the optimal experimental conditions rendered the model non-predictive.

\subsection*{Fundamental Limits to Parameter Estimation in Sloppy Systems}

Finally, we use the concept of a sloppy system introduced above to explore the limits to which parameters can be accurately estimated and the effect on the predictive power of the model.  There are three cases to consider, corresponding to the three scenarios depicted in Figure~\ref{fig:sloppysystemcartoon}.

The first case arises when there is a clear separation between the relevant physics and the irrelevant mechanisms.  In this case, a complete model would have two well-separated groups of eigenvalues, similar to case 1 in Figure~\ref{fig:sloppysystemcartoon}.  In this case, the irrelevant mechanisms can be safely ignored and the remaining parameters can be estimated to more-or-less arbitrary accuracy.  As mentioned in the introduction, these cases can often be explained by a  ``small'' parameter in the system, such as a ratio of well-separated time or length scales.  In these cases, the small parameter explicitly suppresses the influence of the irrelevant details.  The ideal gas law, for example, gives very accurate predictions for pressure and volume over a wide range of densities and temperatures without accounting for fluctuations due statistical mechanics considerations.

The second case is when the model ignores several relevant details.  If a complete model has large eigenvalues with no analog in an approximate model, then the approximate model will give poor fits to data as we saw for the two test cases above.  In this case, the parameter uncertainty is not the bottleneck to model efficacy and is largely irrelevant.  Rather, the systematic errors in the model lead to inaccurate predictions and the model should be refined.  Our results suggest that this case may easily occur when optimal experimental design is applied to complex models.

Finally, consider the case of a sloppy model for which there is no clear separation between the important and unimportant model details.  For many complex systems this appears to be a common occurrence since the eigenvalues are often uniformly spaced on a log scale. 

In order for an approximate model to fit the data, there must be a parameter in the model that can be identified with each relevant mechanisms in the true model.  Furthermore, in order to have accurate parameter estimates, there should be no small eigenvalues in the approximate model.  Therefore, the ideal case is one in which there is a one-to-one correspondence between the large FIM eigenvalues of the complete model and the parameters of the approximate model.  

For the ideal scenario in which the parameters of the approximate model correspond to the subspace of largest eigenvalues in the complete model, we can estimate the magnitude of the model error.  A missing parameter combination with eigenvalue $\lambda$ will typically contribute an amount $\lambda$ to the sum of squared error.  This is because $\sqrt{\lambda}$ is approximately equal to the width of the ``model manifold'' as in Figure~\ref{fig:RelevantDefine}.  Therefore, the cost (i.e., squared error) of ignoring this parameter will be $\lambda$.  

Because eigenvalues in sloppy models are logarithmically spaced, we assume the approximate model is missing eigenvalues from a geometric series with ratio $r$.  It follows that the contribution to the $\chi^2$ value from model error is $\lambda_0 r/(1 - r)$ where $\lambda_0$ is the smallest eigenvalue in the approximate model:
\begin{equation}
  \label{eq:chisquaredwitherror}
  \chi^2 \approx (M-N) + \frac{\lambda_0 r}{1 - r},
\end{equation}
from which it follows that
\begin{equation}
  \label{eq:festimatelambda}
  f = \sqrt{\frac{\lambda_0 r}{(1-r)(M-N)}}.
\end{equation}
Requiring that $f < 1$ so that the model error is less than the experimental noise gives
\begin{equation}
  \label{eq:lamba0}
  \lambda_0 < \frac{1-r}{r} \left( M - N \right)
\end{equation}
as the threshold below which eigenvalues can be safely ignored by the model.

\section*{Discussion}

\begin{figure}
  \includegraphics[width=5.5in]{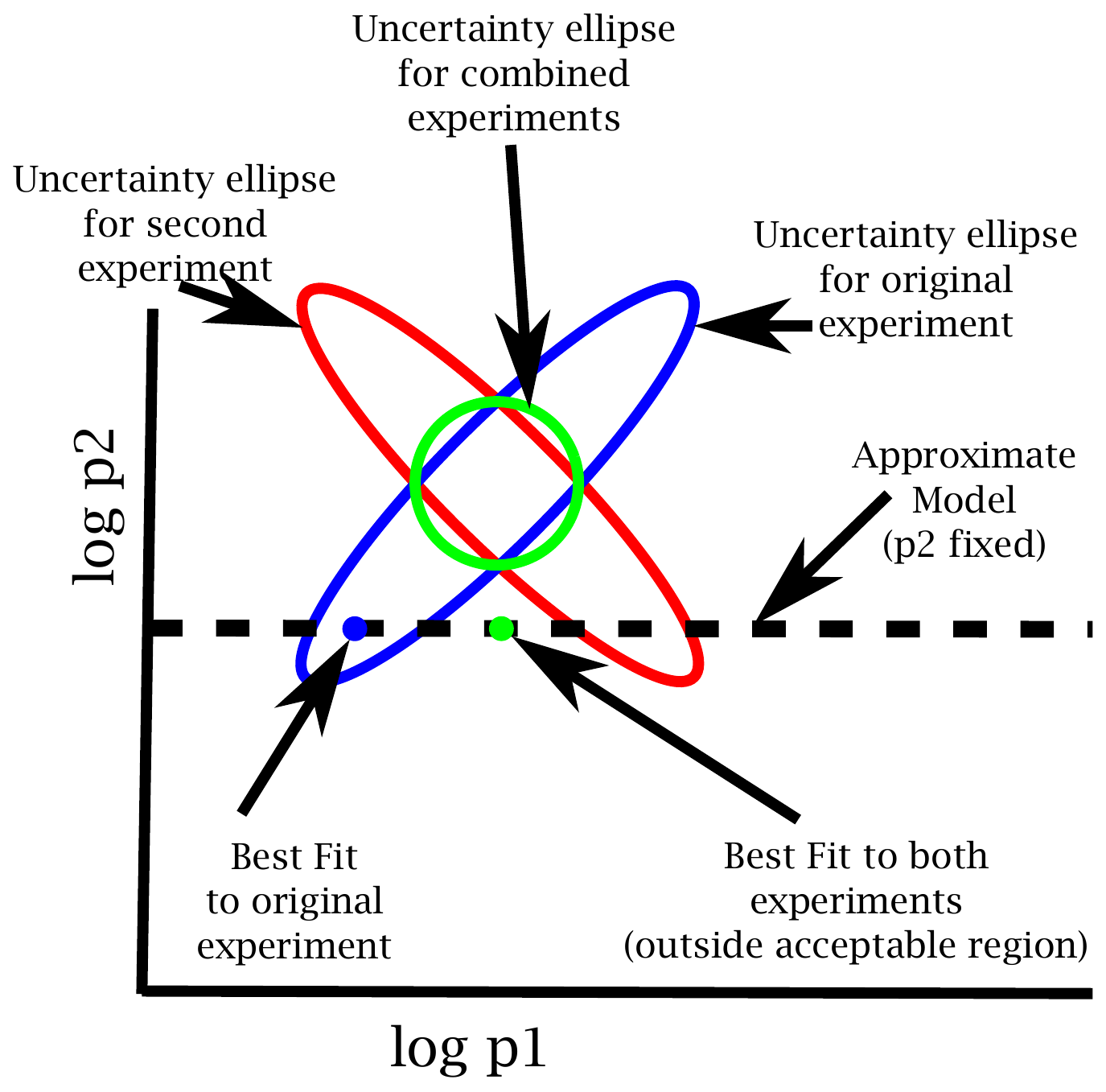}
  \caption{\label{fig:ellipses} \textbf{Uncertainty Ellipses and Approximate Models.} Parameters inside the ellipses are consistent with the data.  Experimental design identifies complementary experiments to minimize the region of consistent parameters.  If the approximate model does not include the this region, the model will be non-predictive for the collection of experiments.}
\end{figure}

In this paper we have explored the relationship among model discrepancy, experimental design, and parameter estimation.  Figure~\ref{fig:ellipses} summarizes are primary result.  When trying to fit complementary experiments to an approximate model, the best fit may often give an inadequate fit to the data.  We explained this result by introducing the concept of \emph{sloppy systems} as a generalization of sloppy models.  Since models are always incomplete, we argued that sloppy models can always be made more accurate by including additional parameters.  In addition to making irrelevant parameters identifiable, optimally chosen experiments may often make the ommitted parameters identifiable too, as illustrated in Figures~\ref{fig:sloppysystemcartoon} and Figure~\ref{fig:eigenvaluesegfr}.  We have demonstrated this for models of EGFR signaling and DNA repair.  We have constructed a simple hyper-model to quantify model error and shown that if a model does not give a good fit then its predictive power is dramatically reduced.  For the two cases considered here, the models are more predictive with unconstrained parameters when fit to a few experiments than they are after fitting to several optimally selected experimental.

\subsection*{Mechanistic Mass-action vs.~Approximate Michaelis-Menten kinetics}

It is perhaps surprising that the approximate Michaelis-Menten model is inadequate even if Eq.~\eqref{eq:MMcond} is satisfied.  However, one should remember that this condition was derived for a single enzyme-substrate reaction in isolation.  One possible explanation of our results is that approximate Michaelis-Menten kinetics are not valid in a network.  This explanation is problematic however, because the approximate Michaelis-Menten model had been used previously to fit real experimental data and make falsifiable predictions for new experiments.  Indeed, in spite of its dubious status, the Michaelis-Menten approximation is often used with much success in many systems biology models.  Therefore, while it is true that the Michaelis-Menten approximation is not \emph{generally} valid, there is considerable evidence that it may sometimes be an safe approximation.

To complicate matters, we have forced Eq.~\eqref{eq:MMcond} to be satisfied by requiring $K_M$ to be very large.  Naively, one would expect this restriction to lead to $K_m$ being unidentifiable.  While this would also be true for measurements of a single reaction, it does not generalize to the network case, as our results demonstrate.  

Furthermore, as the DNA repair results show, our results are not specific to the question of approximate Michaelis-Menten kinetics.  Rather, we have shown that the general question of which physical details are necessary to include in a sloppy model can depend strongly, and in unexpected ways, on which combinations of experiments the model is to explain.  

\subsection*{Implications for Optimal Experimental Design}

A common use for optimal experimental design is model falsification.  Demonstrating the shortcomings of model is hopefully accompanied by new insights into the system's behavior.  Since none of models we have considered here were considered ``correct'' in the reductionist sense, demonstrating that they are incomplete is not profound in itself.  We suggested above that errors in the fit could be used to motivate new hypotheses about microscopic mechanisms.  This possibility is beyond the scope of the current work that focuses on the implications for parameter estimation in sloppy models.

A potential alternative to experimental design for parameter estimation, is experimental design to constrain model predictions\cite{casey2007optimal,transtrum2012optimal}.   Rather than constrain parameter estimates, one seeks to identify a small number of experimental observations that are controlled by the same few parameter combinations as the prediction one would like to make.  In this approach, the model parameters remain sloppy, but the model may be predictive in spite of uncertainty about microscopic details.

It may be surprising that a model may be more predictive in the unidentifiable regime than in the identifiable regime.  The predictivity of the an unidentifiable model is enabled by the narrow widths of the model manifold in Figure~\ref{fig:RelevantDefine}.  The narrow widths guarantee that even infinite fluctuations in parameters do not correspond to large fluctuations in predictions.  It has been suggested elsewhere that ``sloppiness'' can explain why models that make many uncontrolled approximations may be usefully predictive\cite{gutenkunst2007universally,transtrum2015perspective}.  Our results lend some support to this hypothesis; for our test cases removing sloppiness was always accompanied by a decrease in the predictive ability of the model.


\subsection*{Parameter Estimation in Sloppy Models}


In previous work, sloppiness has been viewed as a challenge to be overcome or as a disease to be cured\cite{apgar2010sloppy,tonsing2014cause,chis2014sloppy}.  From this perspective the major challenge of sloppy models has been assumed to be the small eigenvalues of the FIM corresponding to practically unidentifiable parameter combinations.  This in turn has led (incorrectly) to the conflation of sloppiness and practical unidentifiability.  As we have argued here, the near uniform spacing of the eigenvalues (on a log scale) also pose unique challenges for parameter estimation because there is no clear cutoff between relevant and irrelevant mechanisms.

In order for an approximate model to be effective, it is important that the microscopic details ommited from the model be irrelevant, i.e., unidentifiable.   When modeling systems for which all the relevant mechanisms are known, the validity of the model can usually be justified by small parameters, e.g., separated scales or distance from a critical point.  The small parameter guarantees that the FIM eigenvalues for the irrelevant mechanisms are well-separated from those of the relevant mechanisms (e.g., column 3 in Figure~\ref{fig:SloppyvsIdentifiable}).   Some amount of unidentifiability in the physical system is therefore important for effective modeling.  

For many complex systems, no such (known) small parameter exists and sloppy model analysis reveals that there is no sharp distinction between the relevant and irrelevant mechanisms.  We speculate that in many cases the \emph{system} (not just the model) is intrinsically sloppy because there is no intrinsic scale separation to suppress irrelevant mechanisms in the system.  Therefore, a sequence of mechanistically more realistic models would have an eigenvalue structure closer to that in column 1 in Figure 1 rather than Figure 3.  If that is the case, then one should not expect there to exist a mathematical model that can both be accurately calibrated and accurately predict the system behavior.  There will always be several parameters that are marginal, i.e., not tightly constrained by data but are nevertheless necessary to explain the system behavior.  In this case there is a fundamental limit to the efficacy of optimal experimental design: attempting to constrain the marginal parameters of a model of a sloppy system reduces the accuracy of the model and limits its predictive ability as we have seen.

Rather than posing a problem for parameter identification in models of complex systems, we argue here that sloppiness is important for successful modeling.  Sloppy model analysis reveals that in many cases a behavior of interest is controlled by only a small number of parameter combinations.  This observation has been used to explain why relatively simple models can make useful predictions.  Indeed, it has been argued elsewhere that sloppiness may help explain why the world in its microscopic complexity is comprehensible at different scales\cite{transtrum2015perspective}.  Our results give credence to this position since removing sloppiness from a model reduced its predictive ability.

Another approach is to remove the sloppy parameters from the model.  In principle, another simple model may exist whose parameters correspond to the few relevant parameter combinations in the sloppy model.  Parameter estimation in such a model would be relatively straightforward.  Recent advances in model reduction suggest that systematic construction of simple models from complex representations may be generally possible\cite{transtrum2014model,transtrum2015perspective,transtrum2016bridging}.

\subsection*{Relevant and Irrelevant Parameters in Sloppy Systems}

In some branches of physics, the distinction between relevant, irrelevant, and marginal parameters is defined rigorously in terms of the stability of the collective behavior to microscopic variations in mechanistic details as measured by a renormalization group flow.  In that context, relevant parameters correspond to degrees of freedom that must be tuned to achieve a behavior.  In this work, we have used relevant and irrelevant less precisely as synonyms for identifiable and unidentifiable as measured by the FIM eigenvalues.  This equivalence is reasonable because the the identifiable parameters are those that must be tuned to reproduce a behavior.  The equivalence of these definitions  was demonstrated in reference\cite{machta2013parameter}.  However, one of the hallmark features of sloppy models is the roughly uniform spacing of FIM eigenvalues, making it difficult to make a clear delineation between relevant and irrelevant parameters.  

Lacking a clear cutoff between important and unimportant parameter directions means that some physical mechanisms may be either relevant or irrelevant depending on the experimental conditions.  We have shown this explicitly for the two cases considered here.  The model of Brown et al.\cite{brown2003statistical} contained all of the relevant mechanisms (and many irrelevant ones) for explaining several experimental observations of an EGFR pathway.  In contrast, it did not contain all of the relevant mechanisms for experimental conditions proposed by Apgar et al.\cite{apgar2010sloppy}.  Similarly, the LPL model is sufficient for modeling single radiation doses, while the RS model is necessary for modeling sequences of varied radiation doses and neither contains all the relevant mechanisms for modeling the experiments described in this work.  

These results demonstrate the need for a theory of modeling and approximation that identifies which physical mechanisms are relevant for explaining different collective system behaviors.   We have described two approaches that could be the beginnings of such a theory.  First, we have introduced the concept of a sloppy system in which multiple models of varying complexity describe the same observations.  Second, we have used a hyper-model to quantify the limitations of a model.  Although, most of these ideas have existed in some form in the literature, the unique contribution of this work is synthesizing the concepts to explain why sloppy models pose unique challenges for system identification and why these problems are not shared by unidentifiable models that are not sloppy.  

Because simple models are not complete, they cannot make accurate predictions for all experimental conditions.  Of course, it is possible to extend a model by including more details in order to extend its range of validity.  In principle, a single, monolithic model could accurately predict the outcome of all possible experiments.  This possibility underlies the concept of a sloppy system.  Microscopically complete models effectively act as numerical experiments and are a precursor to a more complete theory.  In advancing to a more complete understanding of a system, we believe it is useful to consider multiple models of varying complexity and try to understand their limitations.   Simultaneously considering multiple representations creates a rich and insightful theory into the mechanisms driving behavior that allow for abstraction and generalization.  We believe that accounting for the approximations and context of a model are essential for successful modeling.

\subsection*{Conclusion}

In this paper we have proposed a ``sloppy system hypothesis.''  We speculate that the prevalance of sloppy models in complex biological systems (and other areas of science) is not due to a limitation in the measurement structure of the system, but reflects a property intrinsic to the system itself.  Because many complex systems lack an intrinsic scale separation (i.e., ``small parameter'' as discussed in the introduction), there is no mechanism whereby irrelevant details are necessarily suppressed in the model.  Consequently, corrections to the mathematical model are relevant at all scales so that an accurate model will necessarily have several ``marginal parameters'' as in Figure~\ref{fig:RelevantDefine}.  This hypothesis suggests that there is a fundamental limitation of optimal experimental design in sloppy systems due to these marginal parameters; attempting to constrain the marginal parameters of a model of a sloppy system reduces the accuracy of the model and limits its predictive ability.  We have demonstrated this phenomenon on two complex biological systems, EGFR signaling and DNA repair.  

Mathematical modeling in the face of structural uncertainty is a problem of growing importance across science\cite{smith2013uncertainty,kennedy2001bayesian,qian2006building,apley2006understanding,bayarri2007framework,reichert2009analyzing}.  Because mathematical models by their very nature are not exact replicas of physical processes, it is essential that they include the physical details relevant to the behavior of interest.  In some branches of science, most notably several areas in physics, the equations governing some phenomena are sufficiently well-understood that numerical simulations come very near to being surrogates for real experiments.  When this is the case, accurate parameter estimates reduce uncertainty in the model's predictions.  However, in many areas of complexity science, particularly for systems with fewer symmetries and less homogeneity, which physical details are relevant for explaining a particular behavior remains the theoretical bottleneck.  

Our results suggest that there is a need for better understanding of and accounting for the approximations in complex models.  In particular, optimal experimental design methods should limit their search space to those experiments for which the model is an accurate approximation.  In spite of the growing documentation of microscopic biological mechanisms, it is difficult to predict how the errors introduced by a given approximation (such as the steady state approximation) will propogate to the predictions for the system's collective behavior.  In other words, it is difficult to know a priori which mechanisms are relevant for a particular behavior.  We believe that better quantification of uncertainty will enable improved methods of experimental design and the development of accurate models for predicting behavior in complex systems.

\section*{Materials and Methods}

\subsection*{Ethics Statement}

Animals were maintained in an Association for Assessment and Accreditation of Laboratory Animal Care approved facility, and in accordance with current regulations of the United States Department of Agriculture and Department of Health and Human Services. The experimental protocol was approved by, and in accordance with, institutional guidelines established by the Institutional Animal Care and Use Committee.

UT MD Anderson Cancer Center

ACUF \#:  00001061 RN00

Date Approved:  1/24/2014

Expiration Date:  1/2/2017

\subsection*{EGFR Models and Simulation}

We use the model of EGFR signaling due to Brown et al.\cite{brown2004statistical} formulated in terms of approximate Michaelis-Menten kinetics.  We also constructed a similar model using mechanistic mass action kinetics.  This replaces each approximate Michaelis-Menten reaction with two mass-action steps: $E + S \rightleftharpoons ES \rightarrow E + P$.  We first model each chemical reaction as an enzyme and substrate reversibly binding into an enzyme-substrate complex, and then dissociating to yield the original enzyme and the product.  This gives four nonlinear ordinary differential equations (ODEs) for each enzyme substrate ppreaction, including one each for the changes in concentration of the enzyme, the substrate, the enzyme-substrate complex, and the resulting product. In total, modeling the EGFR network using the same topology as the Brown model by means of mechanistic mass action kinetics requires 54 independent, nonlinear ODES with 70 parameters.  These equations are given in the supplement along with an sbml implementation of the mechanistic model.

All models were simulated using in-house C, FORTRAN and python routines, including methods to automatically calculate parameter sensitivities, included in the supporting information.  We use the approximate Michaelis-Menten model to simulate the original seven laboratory experiments performed by Brown et al.~using parameter values from reference\cite{brown2004statistical}.  We then add random noise to results of this simulation and treat the results as if they were actual laboratory results for the experiments.  Finally, we fit this data to the mechanistic mass-action model using the geodesic Levenberg-Marquardt algorithm\cite{transtrum2010nonlinear,SFGeoLM}.   In order to help avoid complications in which the fit is prematurely stuck at manifold boundaries (as described in\cite{transtrum2010nonlinear}), all fits were done with regularizing terms to keep parameters from drifting to infinite values.  We use regularizing terms that take the form $w_i (\log x_i/x_{i0})^2$ for each parameter $x_i$.  Fits were repeated for many different values of $x_{i0}$ (i.e., the point at which the regularization was centered) and weights $w_i$.  We observe that our final fits were robust to these choices.  

Using the mechanistic model with parameters from fitting the Brown experiments, we simulate the five experimental conditions proposed by Apgar et al.\cite{apgar2010sloppy} and add random noise.  We then fit the approximate model to the these data as before.

\subsection*{Radiation Experiments}

Experimental methods are the same as in reference\cite{sheu2013use}.  C3Hf/KamLaw mice were exposed to whole body irradiation using 300 kVp X-rays at a dose rate of 1.84 Gy/min, and the number of viable jejunal crypts was determined using the microcolony assay. 14 Gy total dose was split into unequal first and second fractions separated by 4 h. Data were analyzed using the LQ model, the lethal potentially lethal (LPL) model, and a repair-saturation (RS) model.




\section*{Acknowledgments}

The authors thank Jim Sethna for reading the manuscript and providing helpful feedback.

%
%
%
\bibliography{../../References/References}

\begin{thebibliography}{10}

\bibitem{rosenblueth1945role}
Rosenblueth A, Wiener N.
\newblock The role of models in science.
\newblock Philosophy of science. 1945;12(4):316--321.

\bibitem{goldenfeld1999simple}
Goldenfeld N, Kadanoff LP.
\newblock Simple lessons from complexity.
\newblock Science. 1999;284(5411):87--89.

\bibitem{rothenberg1971identification}
Rothenberg TJ.
\newblock Identification in parametric models.
\newblock Econometrica: Journal of the Econometric Society. 1971;p. 577--591.

\bibitem{cobelli1980parameter}
Cobelli C, Distefano~3rd JJ.
\newblock Parameter and structural identifiability concepts and ambiguities: a
  critical review and analysis.
\newblock American Journal of Physiology-Regulatory, Integrative and
  Comparative Physiology. 1980;239(1):R7--R24.

\bibitem{chis2014sloppy}
Chis OT, Banga JR, Balsa-Canto E.
\newblock Sloppy models can be identifiable.
\newblock arXiv preprint arXiv:14031417. 2014;.

\bibitem{brown2003statistical}
Brown KS, Sethna JP.
\newblock Statistical mechanical approaches to models with many poorly known
  parameters.
\newblock Physical Review E. 2003;68(2):021904.

\bibitem{brown2004statistical}
Brown KS, Hill CC, Calero GA, Myers CR, Lee KH, Sethna JP, et~al.
\newblock The statistical mechanics of complex signaling networks: nerve growth
  factor signaling.
\newblock Physical biology. 2004;1(3):184.

\bibitem{waterfall2006sloppy}
Waterfall JJ, Casey FP, Gutenkunst RN, Brown KS, Myers CR, Brouwer PW, et~al.
\newblock Sloppy-model universality class and the Vandermonde matrix.
\newblock Physical review letters. 2006;97(15):150601.

\bibitem{gutenkunst2007universally}
Gutenkunst RN, Waterfall JJ, Casey FP, Brown KS, Myers CR, Sethna JP.
\newblock Universally sloppy parameter sensitivities in systems biology models.
\newblock PLoS computational biology. 2007;3(10):e189.

\bibitem{daniels2008sloppiness}
Daniels BC, Chen YJ, Sethna JP, Gutenkunst RN, Myers CR.
\newblock Sloppiness, robustness, and evolvability in systems biology.
\newblock Current opinion in biotechnology. 2008;19(4):389--395.

\bibitem{frederiksen2004bayesian}
Frederiksen SL, Jacobsen KW, Brown KS, Sethna JP.
\newblock Bayesian ensemble approach to error estimation of interatomic
  potentials.
\newblock Physical review letters. 2004;93(16):165501.

\bibitem{machta2013parameter}
Machta BB, Chachra R, Transtrum MK, Sethna JP.
\newblock Parameter space compression underlies emergent theories and
  predictive models.
\newblock Science. 2013;342(6158):604--607.

\bibitem{apgar2010sloppy}
Apgar JF, Witmer DK, White FM, Tidor B.
\newblock Sloppy models, parameter uncertainty, and the role of experimental
  design.
\newblock Molecular BioSystems. 2010;6(10):1890--1900.

\bibitem{chachra2011comment}
Chachra R, Transtrum MK, Sethna JP.
\newblock Comment on ``Sloppy models, parameter uncertainty, and the role of
  experimental design''.
\newblock Molecular BioSystems. 2011;7(8):2522--2522.

\bibitem{tonsing2014cause}
T{\"o}nsing C, Timmer J, Kreutz C.
\newblock Cause and cure of sloppiness in ordinary differential equation
  models.
\newblock Physical Review E. 2014;90(2):023303.

\bibitem{chachra2012structural}
Chachra R, Transtrum MK, Sethna JP.
\newblock Structural susceptibility and separation of time scales in the van
  der Pol oscillator.
\newblock Physical Review E. 2012;86(2):026712.

\bibitem{transtrum2014model}
Transtrum MK, Qiu P.
\newblock Model Reduction by Manifold Boundaries.
\newblock Physical Review Letters. 2014;113(9):098701.

\bibitem{transtrum2016bridging}
Transtrum MK, Qiu P.
\newblock Bridging Mechanistic and Phenomenological Models of Complex
  Biological Systems.
\newblock PLoS Computational Biology. 2016;8:e1004915.

\bibitem{gutenkunst2008sloppiness}
Gutenkunst RN.
\newblock Sloppiness, Modeling, and Evolution in Biochemical Networks.
\newblock Cornell University; 2008.

\bibitem{transtrum2015perspective}
Transtrum MK, Machta BB, Brown KS, Daniels BC, Myers CR, Sethna JP.
\newblock Perspective: Sloppiness and emergent theories in physics, biology,
  and beyond.
\newblock The Journal of Chemical Physics. 2015;143(1):010901.

\bibitem{transtrum2010nonlinear}
Transtrum MK, Machta BB, Sethna JP.
\newblock Why are nonlinear fits to data so challenging?
\newblock Physical review letters. 2010;104(6):060201.

\bibitem{murray1993differential}
Murray MK, Rice JW.
\newblock Differential geometry and statistics. vol.~48.
\newblock CRC Press; 1993.

\bibitem{amari2007methods}
Amari Si, Nagaoka H.
\newblock Methods of information geometry. vol. 191.
\newblock American Mathematical Soc.; 2007.

\bibitem{transtrum2011geometry}
Transtrum MK, Machta BB, Sethna JP.
\newblock Geometry of nonlinear least squares with applications to sloppy
  models and optimization.
\newblock Physical Review E. 2011;83(3):036701.

\bibitem{pukelsheim1993optimal}
Pukelsheim F.
\newblock Optimal design of experiments. vol.~50.
\newblock siam; 1993.

\bibitem{faller2003simulation}
Faller D, Klingm{\"u}ller U, Timmer J.
\newblock Simulation methods for optimal experimental design in systems
  biology.
\newblock Simulation. 2003;79(12):717--725.

\bibitem{cho2003experimental}
Cho KH, Shin SY, Kolch W, Wolkenhauer O.
\newblock Experimental design in systems biology, based on parameter
  sensitivity analysis using a monte carlo method: A case study for the
  tnf$\alpha$-mediated nf-$\kappa$ b signal transduction pathway.
\newblock Simulation. 2003;79(12):726--739.

\bibitem{casey2007optimal}
Casey FP, Baird D, Feng Q, Gutenkunst RN, Waterfall JJ, Myers CR, et~al.
\newblock Optimal experimental design in an epidermal growth factor receptor
  signalling and down-regulation model.
\newblock IET systems biology. 2007;1(3):190--202.

\bibitem{balsa2008computational}
Balsa-Canto E, Alonso AA, Banga JR.
\newblock Computational procedures for optimal experimental design in
  biological systems.
\newblock IET systems biology. 2008;2(4):163--172.

\bibitem{apgar2008stimulus}
Apgar JF, Toettcher JE, Endy D, White FM, Tidor B.
\newblock Stimulus design for model selection and validation in cell signaling.
\newblock PLoS computational biology. 2008;4(2):e30.

\bibitem{erguler2011practical}
Erguler K, Stumpf MP.
\newblock Practical limits for reverse engineering of dynamical systems: a
  statistical analysis of sensitivity and parameter inferability in systems
  biology models.
\newblock Molecular BioSystems. 2011;7(5):1593--1602.

\bibitem{transtrum2012optimal}
Transtrum MK, Qiu P.
\newblock Optimal experiment selection for parameter estimation in biological
  differential equation models.
\newblock BMC bioinformatics. 2012;13(1):181.

\bibitem{chung2012experimental}
Chung M, Haber E.
\newblock Experimental design for biological systems.
\newblock SIAM Journal on Control and Optimization. 2012;50(1):471--489.

\bibitem{mdluli2015efficient}
Mdluli T, Buzzard GT, Rundell AE.
\newblock Efficient Optimization of Stimuli for Model-Based Design of
  Experiments to Resolve Dynamical Uncertainty.
\newblock PLoS Comput Biol. 2015;11(9):e1004488.

\bibitem{he2010maximin}
He F, Brown M, Yue H.
\newblock Maximin and Bayesian robust experimental design for measurement set
  selection in modelling biochemical regulatory systems.
\newblock International Journal of Robust and Nonlinear Control.
  2010;20(9):1059--1078.

\bibitem{bazil2012global}
Bazil JN, Buzzard GT, Rundell AE.
\newblock A global parallel model based design of experiments method to
  minimize model output uncertainty.
\newblock Bulletin of mathematical biology. 2012;74(3):688--716.

\bibitem{vanlier2012bayesian}
Vanlier J, Tiemann CA, Hilbers PA, van Riel NA.
\newblock A Bayesian approach to targeted experiment design.
\newblock Bioinformatics. 2012;28(8):1136--1142.

\bibitem{weber2012trajectory}
Weber P, Kramer A, Dingler C, Radde N.
\newblock Trajectory-oriented bayesian experiment design versus Fisher
  A-optimal design: an in depth comparison study.
\newblock Bioinformatics. 2012;28(18):i535--i541.

\bibitem{busetto2013near}
Busetto AG, Hauser A, Krummenacher G, Sunn{\aa}ker M, Dimopoulos S, Ong CS,
  et~al.
\newblock Near-optimal experimental design for model selection in systems
  biology.
\newblock Bioinformatics. 2013;29(20):2625--2632.

\bibitem{liepe2013maximizing}
Liepe J, Filippi S, Komorowski M, Stumpf MP.
\newblock Maximizing the information content of experiments in systems biology.
\newblock PLoS Comput Biol. 2013;9(1):e1002888.

\bibitem{pauwels2014bayesian}
Pauwels E, Lajaunie C, Vert JP.
\newblock A bayesian active learning strategy for sequential experimental
  design in systems biology.
\newblock BMC Systems Biology. 2014;8(1):102.

\bibitem{oda2005comprehensive}
Oda K, Matsuoka Y, Funahashi A, Kitano H.
\newblock A comprehensive pathway map of epidermal growth factor receptor
  signaling.
\newblock Molecular systems biology. 2005;1(1).

\bibitem{segel1989quasi}
Segel LA, Slemrod M.
\newblock The quasi-steady-state assumption: a case study in perturbation.
\newblock SIAM review. 1989;31(3):446--477.

\bibitem{smith2013uncertainty}
Smith RC.
\newblock Uncertainty Quantification: Theory, Implementation, and Applications.
  vol.~12.
\newblock SIAM; 2013.

\bibitem{kennedy2001bayesian}
Kennedy MC, O'Hagan A.
\newblock Bayesian calibration of computer models.
\newblock Journal of the Royal Statistical Society Series B, Statistical
  Methodology. 2001;p. 425--464.

\bibitem{qian2006building}
Qian Z, Seepersad CC, Joseph VR, Allen JK, Wu CJ.
\newblock Building surrogate models based on detailed and approximate
  simulations.
\newblock Journal of Mechanical Design. 2006;128(4):668--677.

\bibitem{apley2006understanding}
Apley DW, Liu J, Chen W.
\newblock Understanding the effects of model uncertainty in robust design with
  computer experiments.
\newblock Journal of Mechanical Design. 2006;128(4):945--958.

\bibitem{bayarri2007framework}
Bayarri MJ, Berger JO, Paulo R, Sacks J, Cafeo JA, Cavendish J, et~al.
\newblock A framework for validation of computer models.
\newblock Technometrics. 2007;49(2).

\bibitem{reichert2009analyzing}
Reichert P, Mieleitner J.
\newblock Analyzing input and structural uncertainty of nonlinear dynamic
  models with stochastic, time-dependent parameters.
\newblock Water Resources Research. 2009;45(10).

\bibitem{kuepfer2007ensemble}
Kuepfer L, Peter M, Sauer U, Stelling J.
\newblock Ensemble modeling for analysis of cell signaling dynamics.
\newblock Nature biotechnology. 2007;25(9):1001--1006.

\bibitem{feig2004mmtsb}
Feig M, Karanicolas J, Brooks CL.
\newblock MMTSB Tool Set: enhanced sampling and multiscale modeling methods for
  applications in structural biology.
\newblock Journal of Molecular Graphics and Modelling. 2004;22(5):377--395.

\bibitem{ayton2007multiscale}
Ayton GS, Noid WG, Voth GA.
\newblock Multiscale modeling of biomolecular systems: in serial and in
  parallel.
\newblock Current opinion in structural biology. 2007;17(2):192--198.

\bibitem{murtola2009multiscale}
Murtola T, Bunker A, Vattulainen I, Deserno M, Karttunen M.
\newblock Multiscale modeling of emergent materials: biological and soft
  matter.
\newblock Physical Chemistry Chemical Physics. 2009;11(12):1869--1892.

\bibitem{transtrum2014information}
Transtrum MK, Hart G, Qiu P.
\newblock Information topology identifies emergent model classes.
\newblock arXiv preprint arXiv:14096203. 2014;.

\bibitem{sheu2013use}
Sheu T, Molkentine J, Transtrum MK, Buchholz TA, Withers HR, Thames HD, et~al.
\newblock Use of the LQ model with large fraction sizes results in
  underestimation of isoeffect doses.
\newblock Radiotherapy and Oncology. 2013;109(1):21--25.

\bibitem{thames1985incomplete}
Thames HD.
\newblock An'incomplete-repair'model for survival after fractionated and
  continuous irradiations.
\newblock International Journal of Radiation Biology. 1985;47(3):319--339.

\bibitem{curtis1986lethal}
Curtis SB.
\newblock Lethal and potentially lethal lesions induced by radiation---a
  unified repair model.
\newblock Radiation research. 1986;106(2):252--270.

\bibitem{kiefer1988quantitative}
Kiefer J.
\newblock Quantitative mathematical models in radiation biology.
\newblock Radiation and Environmental Biophysics. 1988;27(3):219--232.

\bibitem{ang1992impact}
Ang K, Jiang G, Guttenberger R, Thames H, Stephens L, Smith C, et~al.
\newblock Impact of spinal cord repair kinetics on the practice of altered
  fractionation schedules.
\newblock Radiotherapy and Oncology. 1992;25(4):287--294.

\bibitem{SFGeoLM}
Transtrum MK. Geodesic {L}evenberg-{M}arquardt Source Code; 2012.
\newblock {h}ttp://sourceforge.net/projects/geodesiclm/.

\end{thebibliography}



\end{document}